\pdfoutput=1

\PassOptionsToPackage{prologue,dvipsnames}{xcolor}
\PassOptionsToPackage{sort}{natbib}

\documentclass[11pt]{article}

\usepackage[final]{acl}

\usepackage{times}
\usepackage{latexsym}

\usepackage[T1]{fontenc}

\usepackage[utf8]{inputenc}

\usepackage{microtype}

\usepackage{inconsolata}

\usepackage{graphicx}

\usepackage{makecell}
\usepackage{multirow}
\usepackage{multicol}
\usepackage{caption}
\usepackage{subcaption}
\usepackage{wrapfig}
\usepackage{tablefootnote}
\usepackage{url}            %
\usepackage{xurl} 
\usepackage{booktabs}       %
\usepackage{amsfonts}       %
\usepackage{nicefrac}       %
\usepackage{microtype}      %
\usepackage{tabularx}
\usepackage[export]{adjustbox}
\usepackage[many]{tcolorbox}

\usepackage{amsmath,bm}
\usepackage{amsfonts}
\usepackage{comment}
\usepackage{amssymb}
\usepackage{mathtools}
\usepackage{bbm}
\usepackage{enumitem}
\usepackage{lipsum}  

\usepackage{pifont}%
\newcommand{\cmark}{{\ding{51}}}%
\newcommand{\xmark}{{\ding{55}}}%
\definecolor{MycolorRed}{HTML}{c00000}
\definecolor{MycolorBlue}{HTML}{2e76b6}
\definecolor{MycolorGreen}{HTML}{71ad47}

\newcommand{\highlight}[2]{\colorbox{#1}{\rule[-0.45ex]{-1pt}{0.90em}#2\hspace*{-1pt}}}
\usepackage{tikz}

\usepackage{soul}
\sethlcolor{yellow}

\makeatletter
\newcommand*{\da@rightarrow}{\mathchar"0\hexnumber@\symAMSa 4B }
\newcommand*{\da@leftarrow}{\mathchar"0\hexnumber@\symAMSa 4C }
\newcommand*{\xdashrightarrow}[2][]{%
  \mathrel{%
    \mathpalette{\da@xarrow{#1}{#2}{}\da@rightarrow{\,}{}}{}%
  }%
}
\newcommand{\xdashleftarrow}[2][]{%
  \mathrel{%
    \mathpalette{\da@xarrow{#1}{#2}\da@leftarrow{}{}{\,}}{}%
  }%
}
\newcommand*{\da@xarrow}[7]{%
  \sbox0{$\ifx#7\scriptstyle\scriptscriptstyle\else\scriptstyle\fi#5#1#6\m@th$}%
  \sbox2{$\ifx#7\scriptstyle\scriptscriptstyle\else\scriptstyle\fi#5#2#6\m@th$}%
  \sbox4{$#7\dabar@\m@th$}%
  \dimen@=\wd0 %
  \ifdim\wd2 >\dimen@
    \dimen@=\wd2 %
  \fi
  \count@=2 %
  \def\da@bars{\dabar@\dabar@}%
  \@whiledim\count@\wd4<\dimen@\do{%
    \advance\count@\@ne
    \expandafter\def\expandafter\da@bars\expandafter{%
      \da@bars
      \dabar@ 
    }%
  }%
  \mathrel{#3}%
  \mathrel{%
    \mathop{\da@bars}\limits
    \ifx\\#1\\%
    \else
      _{\copy0}%
    \fi
    \ifx\\#2\\%
    \else
      ^{\copy2}%
    \fi
  }%
  \mathrel{#4}%
}
\makeatother

\RequirePackage{xspace}
\makeatletter
\DeclareRobustCommand\onedot{\futurelet\@let@token\@onedot}
\def\@onedot{\ifx\@let@token.\else.\null\fi\xspace}

\def\eg{\emph{e.g}\onedot} 

\def\ie{\emph{i.e}\onedot}

\makeatother

\title{Two Heads Are Better Than One: Audio-Visual Speech\\Error Correction with Dual Hypotheses}

\author{
 \textbf{Sungnyun Kim\textsuperscript{1}}\thanks{Equal contribution},
 \textbf{Kangwook Jang\textsuperscript{2}}\textcolor{red}{\footnotemark[1]},
 \textbf{Sungwoo Cho\textsuperscript{1}},
\\
 \textbf{Joon Son Chung\textsuperscript{2}},
 \textbf{Hoirin Kim\textsuperscript{2}},
 \textbf{Se-Young Yun\textsuperscript{1}}
\\ [0.5em]
 \textsuperscript{1}Kim Jaechul Graduate School of AI, KAIST \\
 \textsuperscript{2}School of Electrical Engineering, KAIST
\\ [0.5em]
\texttt{\{ksn4397, dnrrkdwkd12, peter8526,} \\
\texttt{joonson, hoirkim, yunseyoung\}@kaist.ac.kr}
\\
}

\begin{document}
\maketitle

\begin{abstract}
This paper introduces a new paradigm for generative error correction (GER) framework in audio-visual speech recognition (AVSR) that reasons over modality-specific evidences directly in the language space. Our framework, \textbf{DualHyp}, empowers a large language model (LLM) to compose independent $N$-best hypotheses from separate automatic speech recognition (ASR) and visual speech recognition (VSR) models. To maximize the effectiveness of DualHyp, we further introduce \textbf{RelPrompt}, a noise-aware guidance mechanism that provides modality-grounded prompts to the LLM. RelPrompt offers the temporal reliability of each modality stream, guiding the model to dynamically switch its focus between ASR and VSR hypotheses for an accurate correction. Under various corruption scenarios, our framework attains up to 57.7\% error rate gain on the LRS2 benchmark over standard ASR baseline, contrary to single-stream GER approaches that achieve only 10\% gain. To facilitate research within our DualHyp framework, we release the code and the dataset comprising ASR and VSR hypotheses at \url{https://github.com/sungnyun/dualhyp}.
\end{abstract}

\section{Introduction}

Recent advancements have introduced GER frameworks that utilize LLMs to refine ASR outputs.
Following the release of $N$-best ASR hypotheses dataset \cite{chen2023hyporadise}, numerous studies demonstrated the efficacy of LLMs in correcting transcriptions based on the hypotheses list \cite{hu2024robustger, hu2024clozeger, mu2024mmger, mu2025mixture}.
These powerful correction frameworks, however, presents a fundamental limitation.
While the performance of the underlying ASR systems is remarkable in controlled environments~\cite{graves2012sequence, zhang2020transformer, chiu2022self, peng2024owsm}, it degrades significantly in noisy real-world conditions where acoustic distortions are prevalent.
To mitigate this challenge, AVSR systems have been developed~\cite{kim2024learning, kim2025mohave, han2024xlavs, shi2022learning, chen2023leveraging}, leveraging complementary visual cues (\eg, lip movements) to enhance robustness against noise.

In the realm of AVSR, integrating visual information into GER frameworks remains a nascent area of research.
Existing methods often employ visual adapters~\cite{ghosh2024lipger} or unified AVSR models~\cite{liu2025avger}, both of which process visual data in the feature space.
This feature-level fusion struggles when audio and visual streams are corrupted independently, as noise from one modality can easily contaminate the unified representation \cite{kim2025multi}.
Moreover, these frameworks heavily rely on a single set of hypotheses generated from one, often error-prone, recognition model. %

To address these limitations, we propose \textbf{DualHyp}, the first GER framework that explicitly maintains modality-specific pathways from separate ASR and VSR systems\,(\S\ref{sec:dualhyp}).
LLM intelligently composes these \textbf{dual-stream hypotheses}, leveraging the model's deep contextual understanding in the \textit{language space} rather than forcing the model to interpret complex audio or video embedding subspaces.
Building upon this, we introduce \textbf{RelPrompt}, a \textbf{noise-aware guidance} mechanism that directs the underlying quality of each modality (\S\ref{sec:relprompt}).
Since LLMs for GER primarily operate within the language space, they lack modality-level grounding and may incorrectly prioritize unreliable sources.
To mitigate this, we incorporate reliability predictors to assess the quality of audio and visual streams, which are fed to the LLM to better elicit the compositional capacity of DualHyp.

Our experiments\,(\S\ref{sec:experiments}) show that this DualHyp approach with RelPrompt significantly outperforms prior single-stream GER frameworks across various audio-visual corruption scenarios.
We also demonstrate its multilingual capabilities as well as improved reasoning with larger LLMs.
Through qualitative analysis\,(\S\ref{sec:analysis}), we investigate the correction mechanism that makes our framework more effective.

\begin{figure*}[!t]
    \centering
    \includegraphics[width=\linewidth]{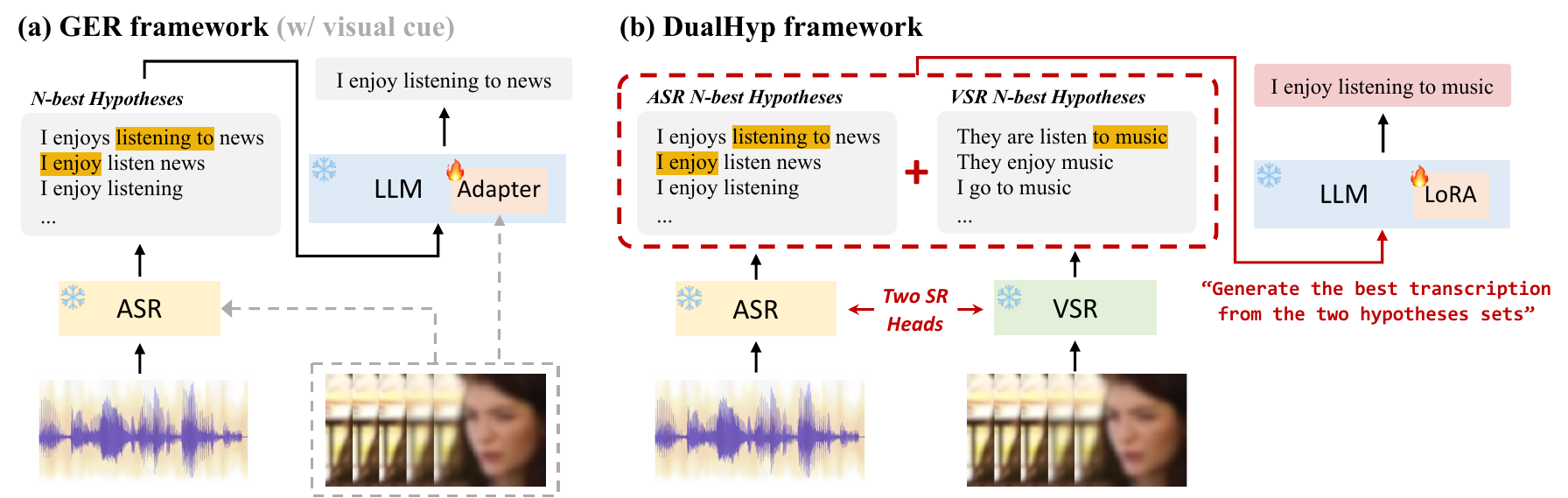}
    \vspace{-18pt}
    \caption{
    (a) Conventional GER frameworks use a single set of ASR hypotheses and (optionally) injects visual features via an adapter or a multimodal encoder. 
    (b) Our DualHyp framework maintains modality separation, using both ASR and VSR heads to generate two distinct sets of textual hypotheses. The LLM performs compositional reasoning on dual hypotheses in the language space to produce a more robust and accurate transcription.
    }
    \label{fig:overview}
\end{figure*}
\section{Related Works}

\noindent
\textbf{Generative error correction for speech.}~~Recently, there has been growing interest in using LLMs for post-hoc correction of speech recognition outputs.
Initial work in GER for ASR demonstrates that LLMs can effectively regenerate transcriptions from $N$-best hypothesis lists~\cite{chen2023hyporadise}. %
Subsequent research has refined this paradigm by exploring novel prompting strategies like cloze-style completion~\cite{hu2024clozeger} or by re-injecting acoustic features to better ground the LLM's corrections~\cite{chen2024uadf, radhakrishnan2023whispering, mu2024mmger, mu2025mixture, liu2025denoising}. These foundational works, however, focus exclusively on correcting hypotheses generated from a single, audio-only stream.

\vspace*{4pt}
\noindent
\textbf{Modality fusion in GER for AVSR.}~~Extending GER to the audio-visual domain presents the central challenge of how to effectively fuse multimodal information. Existing approaches perform this fusion in the feature space, before the final language generation step.
\citet{ghosh2024lipger} involved visual adapters~\citep{houlsby2019parameter, zhang2024llama} to inject lip-reading features directly into the LLM, while \citet{liu2025avger} used dedicated multimodal encoders to create a unified audio-visual representation.
While these methods show promise, their reliance on early, feature-level fusion makes them vulnerable to cross-modal contamination~\cite{hong2022visual}, where corruption in one modality can degrade the quality of the fused representation.

Motivated by prior works highlighting the benefits of modality-specific processing for robustness~\cite{wang2024mlca, kim2025multi, liu2021audio}, our approach is designed to isolate corruptions specific to each modality before error correction.
In contrast to feature-level fusion methods, we achieve this by deliberately delaying the modality fusion to the generation stage where the LLM operates on independent textual hypotheses from separate ASR and VSR models.

\vspace*{4pt}
\noindent
\textbf{End-to-end LLM-based AVSR.}~~It is important to distinguish our GER framework from an orthogonal line of research that uses LLMs for end-to-end\,(E2E) ASR~\cite{ma2024embarrassingly, yu2024connecting, fathullah2024prompting} and AVSR~\cite{cappellazzo2025large, cappellazzo2025adaptive, cappellazzo2025scaling, yeo2025mms, yeo2024visual}. In that paradigm, encoded audio and visual features serve as direct, multimodal prompts for a single generative model.
While promising, our decoupled approach offers significant advantages in flexibility. 

First, our framework is highly modular and can readily use off-the-shelf ASR systems and LLMs.
This contrasts with monolithic E2E models, which require costly pretraining of the entire system for any component update.
Second, the system can be easily improved by refining text-based prompts. This avoids the inherent complexity of designing and aligning cross-modal prompts, which is a central challenge in E2E systems.

\section{DualHyp Framework}
\label{sec:dualhyp}

\subsection{Uni-modal Generative Error Correction}
\label{subsec:unimodal_ger}
Recent works have successfully employed LLMs for GER~\cite{chen2023hyporadise, hu2024robustger, ghosh2024lipger}, where they aim to refine outputs of a uni-modal ASR system.
Given an input utterance, the ASR model first generates an $N$-best list of candidate transcriptions by beam search decoding, denoted as $\mathcal{H}^{\text{asr}}=\{(h_i^a, s_i^a)\}_{i=1}^N$, where $h_i^a$ is the $i$-th hypothesis and $s_i^a$ is the corresponding log-likelihood score. The LLM takes this hypothesis set as an input and generates a corrected transcription $\hat{y}$ via conditional generation:
\begin{equation}
    \hat{y} = \arg\max_{y} P(y \mid \mathcal{H}^{\text{asr}}; {\theta}_\text{LLM}).
\end{equation}

This approach has proven effective in clean acoustic conditions; however, its performance is fundamentally capped by the quality of the initial ASR hypotheses. When the source audio is severely corrupted by noise such as negative signal-to-noise (SNR) level, the resulting hypotheses are too erroneous to provide useful signal for correction, creating a performance bottleneck.

In contrast, visual information such as lip movements offers a complementary modality that is invariant to acoustic noise.
Visual modality has been shown to be particularly useful in disambiguating homophones or recovering missing segments in noisy environments~\cite{kim2022distinguishing, kim2024learning}.
Motivated by this, we propose to extend GER beyond a single-stream hypothesis by incorporating both audio and visual modalities in a unified framework.

\subsection{Oracle Error Analysis}
\label{subsec:oracle_error_analysis}
To ascertain the potential benefits of incorporating a second modality, we conduct an oracle error analysis of speech recognition systems, in addition to standard 1-best word error rate (WER).
This oracle analysis establishes theoretical lower bounds of ASR and VSR systems in two manners~\cite{chen2023hyporadise}: \textit{N-best oracle} ($o_{nb}$), which selects the single best hypothesis from an $N$-best list, and \textit{compositional oracle} ($o_{cp}$), which constructs an optimal transcript by combining correct words from all $N$-best hypotheses.
Table\,\ref{tab:prelim} summarizes 1-best WERs of three speech recognition heads: Whisper-large-v3 \cite{radford2023robust} for audio-only, BRAVEn-large \cite{haliassos2024braven} for visual-only, and Auto-AVSR \cite{ma2023auto} for audio-visual. Whisper attains 25.8\% WER, while Auto-AVSR is slightly stronger (24.9\%), and BRAVEn is markedly weaker (39.7\%), confirming that VSR alone lags in overall accuracy.

The oracle WER results reveal the limitation of single-stream systems and the compelling potential of dual-stream approaches (A\,+\,AV or A\,+\,V). 
While strong individual models like Whisper and Auto-AVSR perform $o_{cp}$ WERs of 13.7\% and 13.6\%, respectively, combining hypotheses from independent audio and visual heads drastically reduces potential errors: \textbf{Whisper\,(ASR)\,+\,BRAVEn\,(VSR)} plummets to 4.5\%.
This gap indicates that audio and video can provide distinct evidence with highly complementary information.
Consequently, an ideal GER model that can compose across ASR and VSR hypotheses could significantly reduce errors relative to single-stream systems.

\subsection{DualHyp: Dual-Stream Hypotheses}
\label{subsec:dualhyp}
Existing GER approaches for AVSR either inject visual data into LLMs via adapters~\cite{ghosh2024lipger} or rely on multimodal encoders that perform early fusion of the modalities~\cite{liu2025avger}. Both strategies have notable drawbacks; feature adaptation is insufficient for transferring rich visual cues, whereas early fusion is susceptible to cross-modal interference or modality bias.

Our approach is guided by a different principle, underscored by perceptual phenomena that the premature fusion of conflicting audio-visual signals can distort recognition outcomes~\cite{mcgurk1976hearing}. Inspired by prior work that embeds audio noise into the \textit{language space}~\cite{hu2024robustger}, we suggest that modality-specific information should be explicitly represented in the language space. This allows the LLM to resolve inconsistencies and compose information from both streams without entangling the signals during the upstream feature processing.

\begin{table}[!t]
\centering
\small
\vspace{5pt}
\addtolength{\tabcolsep}{2pt}
\resizebox{\columnwidth}{!}{
\begin{tabular}{lcccc}
    \toprule
    \textbf{SR Head} & \textbf{Input} & \!\!\textbf{1-best}\!\! & $\pmb{o_{nb}}$ & $\pmb{o_{cp}}$ \\
    \midrule
    Whisper-large-v3 & A & 25.8 & 16.7 & 13.7 \\
    BRAVEn-large & V & 39.7  & 27.8 & 24.6\\
    Auto-AVSR & AV & 24.9 & 16.1 & 13.6 \\
    \midrule
    Whisper\,+\,Auto-AVSR & A\,+\,AV & -- & 7.0 & 4.9 \\
    Whisper\,+\,BRAVEn & A\,+\,V & -- & \textbf{6.4} & \textbf{4.5} \\
    \bottomrule
\end{tabular}
}
\vspace{-5pt}
\caption{WER\,(\%) analysis with different speech recognition heads, evaluated on noise-augmented LRS2. The audio stream is augmented with four types of noise, and the video stream is corrupted with four visual degradation types. For details, refer to Appendix\,\ref{sec:corruption_protocol}. $o_{nb}$: \emph{$N$-best oracle}, $o_{cp}$: \emph{compositional oracle}.
}
\label{tab:prelim}
\end{table}

\begin{table*}[!t]
\small
\centering
\addtolength{\tabcolsep}{-3pt}
\setlength{\fboxsep}{1pt} %
\resizebox{\textwidth}{!}{
\begin{tabular}{l|c|l|c}
\toprule
\multicolumn{2}{l|}{\textit{\textbf{Type 1: Multimodal Fragment Composition}}} & \multicolumn{2}{l}{\textit{\textbf{Type 2: Dominant Modality Refinement}}} \\
\midrule
\textbf{Utterance (ASR\,+\,VSR $\rightarrow$ DualHyp)} & \textbf{WER} & \textbf{Utterance (ASR\,+\,VSR $\rightarrow$ DualHyp)} & \textbf{WER} \\
\midrule
\midrule
    \textbf{\underline{ASR 5-best} ($\mathcal{H}^{\text{asr}}$): } & & \textbf{\underline{ASR 5-best} ($\mathcal{H}^{\text{asr}}$):} & \\
    \highlight{yellow!30}{everyone going into the den} has\highlight{Green!10}{ a fresh chance to} talk it around & 35.7 & \texttt{<unk>} & 100.0 \\
    \highlight{yellow!30}{everyone going into the den} is given\highlight{Green!10}{ a fresh chance to} talk it around & 42.9 & thank you & 100.0 \\
    \highlight{yellow!30}{everyone going into the den} gives you\highlight{Green!10}{ a fresh chance to} talk it around & 42.9 & all right & 100.0 \\
    and \highlight{yellow!30}{everyone going into the den} has\highlight{Green!10}{ a fresh chance to} talk around & 35.7 & the president & 100.0 \\
    \highlight{yellow!30}{everyone going into the den} has\highlight{Green!10}{ a fresh chance to} talk to the ground & 42.9 & god bless you & 100.0 \\[5pt]
    \textbf{\underline{VSR 5-best} ($\mathcal{H}^{\text{vsr}}$):} & & \textbf{\underline{VSR 5-best} ($\mathcal{H}^{\text{vsr}}$):} & \\
    but everyone in today \highlight{orange!20}{gets}\highlight{Green!10}{ a fresh chance to}\highlight{purple!15}{ turn things around} & 35.7 & \highlight{yellow!30}{project management}\highlight{orange!20}{ is really}\highlight{Green!10}{ my special} considering & 14.3 \\
    but everyone as i say \highlight{orange!20}{gets}\highlight{Green!10}{ a fresh chance to}\highlight{purple!15}{ turn things around} & 35.7 & \highlight{yellow!30}{project management}\highlight{orange!20}{ is really} by special considering & 28.6 \\
    but everyone on its day \highlight{orange!20}{gets}\highlight{Green!10}{ a fresh chance to}\highlight{purple!15}{ turn things around} & 35.7 & \highlight{yellow!30}{project management}\highlight{orange!20}{ is really}\highlight{Green!10}{ my special}\highlight{purple!15}{ist} theory & 14.3 \\
    but everyone it is a saying \highlight{orange!20}{gets}\highlight{Green!10}{ a fresh chance to}\highlight{purple!15}{ turn things around} & 35.7 & \highlight{yellow!30}{project management} and really\highlight{Green!10}{ my special} considering & 28.6 \\
    but everyone it is the saying \highlight{orange!20}{gets}\highlight{Green!10}{ a fresh chance to}\highlight{purple!15}{ turn things around} & 28.6 & \highlight{yellow!30}{project management}\highlight{orange!20}{ is really}\highlight{Green!10}{ my special} discovery & 28.6 \\
\midrule
\textbf{\underline{DualHyp output} $(\hat{y})$:} & & \textbf{\underline{DualHyp output} $(\hat{y})$:} & \\
\textbf{\highlight{yellow!30}{everyone going into the den}\highlight{orange!20}{ gets}\highlight{Green!10}{ a fresh chance to}\highlight{purple!15}{ turn things around}} & \textbf{14.3} & \textbf{\highlight{yellow!30}{project management}\highlight{orange!20}{ is really}\highlight{Green!10}{ my special}\highlight{purple!15}{ist}\highlight{blue!10}{ area}} & \textbf{0.0} \\
\midrule
\textbf{\underline{Ground-truth:}} & & \textbf{\underline{Ground-truth:}} & \\
but everyone going into the den gets a fresh chance to turn things round & -- & project management is really my specialist area & -- \\
\bottomrule
\end{tabular}
}
\vspace{-5pt}
\caption{Examples of successful correction via DualHyp framework. The upper hypothesis within each 5-best list has a higher log-likelihood score. The colored highlights trace the origin of word fragments in the final DualHyp output, showing how those are sourced from ASR, VSR, or both, with a word being newly generated by the LLM's internal knowledge. \textbf{\textit{Type 1}} demonstrates the model combining complementary pieces from both modalities, and \textbf{\textit{Type 2}} presents the model identifying and correcting the hypothesis from a more reliable modality.}
\label{tab:dualhyp_case}
\vspace{5pt}
\end{table*}

Thus, based on our analysis in Section\,\ref{subsec:oracle_error_analysis}, we propose \textbf{DualHyp}, a novel GER framework that explicitly leverages separate hypotheses streams from both audio and video modalities. Instead of relying on a single recognizer, we utilize independent, pretrained ASR and VSR models to process an audio-visual pair.
Each recognizer head generates a distinct $N$-best list:
\begin{equation*}
\mathcal{H}^{\text{asr}}=\{(h_i^{\text{a}}, s_i^{\text{a}})\}_{i=1}^{N}, 
\quad 
\mathcal{H}^{\text{vsr}}=\{(h_j^{\text{v}}, s_j^{\text{v}})\}_{j=1}^{N}.
\end{equation*}

We then form a combined \emph{dual} hypotheses set, $\mathcal{H}^{\text{dual}}=\mathcal{H}^\text{asr}\cup\mathcal{H}^\text{vsr}$, which preserves the modality-specific information in each hypothesis set. The LLM is conditioned on this enriched set to generate the DualHyp output:

\begin{equation}
\hat{y} = \arg\max_{y} P(y \mid \mathcal{H}^\text{dual}; \theta_\text{LLM}).
\end{equation}

By maintaining separate modality pathways into the language space, this approach avoids the cross-modal contamination issues seen in early-fusion models. It instead enables the LLM to act as an in-context compositional reasoner~\citep{qiu2022evaluating, an2023context}, cross-referencing the audio and visual evidence to resolve ambiguities and reconstruct the intended utterance. Figure~\ref{fig:overview} illustrates the overview of our DualHyp framework, compared to existing GER approaches.

\paragraph{Analysis.} Table~\ref{tab:dualhyp_case} provides qualitative analysis of DualHyp to show its effectiveness.
We highlight two primary correction mechanisms that explicitly exploit the LLM's capacity for semantic and linguistic reasoning within the language space. \textit{(Type 1) Multimodal Fragment Composition:} The model constructs the output by conducting a word-level semantic alignment across the dual hypotheses. It actively recovers the correct transcription by extracting complementary linguistic fragments, such as preserving the acoustically clear \texttt{den} from the ASR stream while adopting the visually distinct phrase \texttt{turn things around} from the VSR stream. %
\textit{(Type 2) Dominant Modality Refinement:} When one modality provides highly degraded information, the model identifies the more coherent stream and exclusively grounds its refinement process on the dominant modality. %
Furthermore, this refinement process retains the LLM's prior knowledge, as it generates a word not present in any source hypothesis, \ie, \texttt{area}. We provide more examples, including failure cases, in Appendix \ref{app: qualitative_examples}.

\section{Noise-Aware Guidance of DualHyp}
\label{sec:relprompt}

The DualHyp framework enables an LLM to compose information from separate ASR and VSR hypotheses.
However, since LLM operates purely on these text inputs, the model lacks explicit information about the source signal quality, creating a risk of leveraging unreliable, inaccurate hypotheses \cite{hong2023watch}.
To bridge this gap from an LLM perspective, we introduce \textbf{RelPrompt}, a noise-aware guidance mechanism that explicitly informs the LLM about the temporal reliability of each stream.
RelPrompt is achieved by (1) predicting reliability tokens for each modality using external predictors, which are then (2) provided to the LLM's prompt to serve as temporal guidance.

\subsection{Reliability Mask Prediction}

To generate a compact, time-aligned reliability signal, we segment both the audio and video streams to approximate the duration of a single spoken word.
Grounded in the average native English speaking rate~\cite{yuan2006towards, becker2022horse}, we set the chunk size to 0.4 seconds, \ie, 150 wpm.
We process each modality as follows:
\begin{itemize}[itemsep=0pt, leftmargin=*]
    \vspace{-5pt}
    \item \textbf{Audio stream:} The input audio, sampled at 16kHz, is grouped into segments of 6,400 samples (16,000 samples/sec × 0.4 sec).
    \vspace{-3pt}
    \item \textbf{Video stream:} The input video, processed at 25Hz, is grouped into segments of 10 frames (25Hz × 0.4 sec).
    \vspace{-5pt}
\end{itemize}

We then employ two lightweight predictors consisting of 1D convolutional neural networks (CNN) that operate on the intermediate features extracted from the ASR and VSR encoders, thus avoiding additional feature extraction.
For each segment, the predictors produce a discrete token $m_i\in\{\texttt{Clean}, \texttt{Noisy}$, \texttt{Mixed}\}, forming a reliability mask that indicates the quality of the source signal to the LLM.
The ground-truth reliability is labeled as \texttt{Clean} if <10\% of its frames are corrupted, \texttt{Noisy} if >60\% of its frames are corrupted, and \texttt{Mixed} otherwise.
Each predictor outputs a sequence of these tokens for its respective modality:
\vspace*{-5pt}
\begin{equation*}
\vspace*{-2pt}
\mathbf{m}^{\text{a}} = (m^{\text{a}}_1,\ldots,m^{\text{a}}_K), \quad
\mathbf{m}^{\text{v}} = (m^{\text{v}}_1,\ldots,m^{\text{v}}_K).
\end{equation*}

\subsection{Reliability Guidance}

\begin{figure}[!t]
    \centering
    \begin{tcolorbox}[
        width=\linewidth,
        sharp corners=all,
        colback=gray!10,
        boxrule=0.3mm,
        enhanced,
        boxsep=0.1mm,
        left=2mm,
        right=2mm,
    ]
    \footnotesize

    \texttt{Below are the best-hypothesis transcribed from ASR and VSR. Revise it using the words which are only included into other-hypotheses, and write the response for the true transcription. Refer to the audio and video masks for reliability.} \\

    \texttt{\#\#\# ASR Best-hypothesis:} $\{h^{\text{a}}_1\}$ \\
    \texttt{\#\#\# ASR Other-hypotheses:}
    $\{h^{\text{a}}_2~||~\cdots~||~h^{\text{a}}_{N}\}$ \\
    \textcolor{red}{\texttt{\#\#\# Audio Mask:~[C][N][N][M][C]} $\cdots$} \\

    \texttt{\#\#\# VSR Best-hypothesis:} $\{h^{\text{v}}_1\}$ \\
    \texttt{\#\#\# VSR Other-hypotheses:}
    $\{h^{\text{v}}_2~||~\cdots~||~h^{\text{v}}_{N}\}$ \\
    \textcolor{red}{\texttt{\#\#\# Video Mask:~[C][C][C][N][N]} $\cdots$} \\

    \texttt{\#\#\# Response:}

    \end{tcolorbox}
    \vspace{-12pt}
    \includegraphics[width=\linewidth]{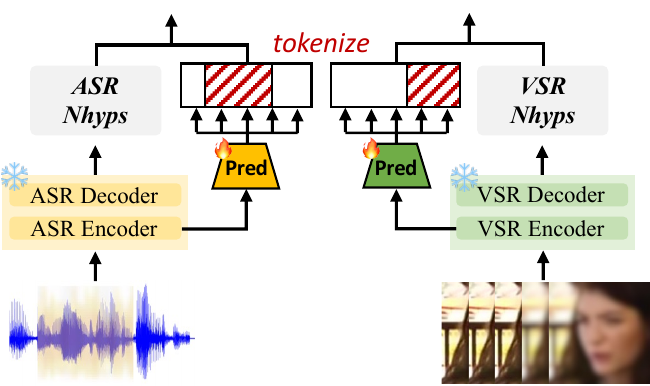}
    \vspace{-18pt}
    \caption{An overview of our DualHyp with RelPrompt. Each predictor uses ASR/VSR encoder features to generate a noise-aware token sequence. These masks accurately guide the LLM to dynamically switch the model's focus between the ASR and VSR hypotheses.}
    \label{fig:overview-2}
\end{figure}

As illustrated in Figure \ref{fig:overview-2}, the reliability token sequences, $\mathbf{m}^\text{a}$ and $\mathbf{m}^\text{v}$ are appended to the dual hypotheses to directly inform the LLM of each modality's temporal reliability.
The entire model is then trained end-to-end, conditioned on both the hypotheses and reliability masks to generate the final transcript:
\vspace*{-5pt}
\begin{equation}
\vspace*{-5pt}
\hat{y}=\arg\max_{y} P\!\left(y \mid \mathcal{H}^{\text{dual}}, \mathbf{m}^{\text{a}}, \mathbf{m}^{\text{v}}; \theta_{\text{LLM}}\right).
\end{equation}

This format allows the LLM to learn the correlation between the reliability tokens and hypotheses quality.
Crucially, this approach avoids the need for explicit word-level alignment, which is infeasible for $N$-best lists with variable lengths and erroneous words \citep{qiu2021learning, gekhman2022red}.
Additionally, the RelPrompt token sequence enhances the interpretability of LLM's reasoning, revealing when the model switches its focus between the ASR and VSR hypotheses.

\begin{table*}[!t]
  \centering
  \small
  \resizebox{\textwidth}{!}{
  \begin{tabular}{lc|llll|l}
    \toprule
    \textbf{Method} & \textbf{Input} & \textbf{Babble (B)} & \textbf{Speech (S)} & \textbf{Music (M)} & \textbf{Natural (N)} & \textbf{Overall (O)} \\
    \midrule
    \color{gray}\textit{ASR oracle $o_{nb}$\,/\,$o_{cp}$} & \color{gray}A & \color{gray}30.9 / 26.9 & \color{gray}19.4 / 14.1 & \color{gray}8.0 / 6.6 & \color{gray}8.5 / 7.4 & \color{gray}16.7 / 13.7 \\
    \color{gray}{\textit{ASR\,+\,VSR oracle $o_{nb}$\,/\,$o_{cp}$}} & \color{gray}A\,+\,V & \color{gray}11.7 / 8.8 & \color{gray}6.6 / 4.4 & \color{gray}3.5 / 2.2 & \color{gray}3.6 / 2.7 & \color{gray}6.4 / 4.5 \\
    \midrule
    Whisper-large-v3~\cite{radford2023robust} & A & 40.0 & 36.5 & 12.7 & 14.2 & 25.8 \\
    BRAVEn-large \cite{haliassos2024braven} & V & - & - & - & - & 39.7$_{(+53.9\%)}$ \\
    \midrule
    GER~\cite{chen2023hyporadise} & A & 39.3$_{(-1.8\%)}$ & 34.4$_{(-5.8\%)}$ & 11.5$_{(-9.4\%)}$ & 13.2$_{(-7.0\%)}$ & 24.6$_{(-4.7\%)}$ \\
    RobustGER~\cite{hu2024robustger} & A & 39.3$_{(-1.8\%)}$ & 33.8$_{(-7.4\%)}$ & 11.7$_{(-7.9\%)}$ & 13.1$_{(-7.7\%)}$ & 24.5$_{(-5.0\%)}$ \\
    LipGER~\cite{ghosh2024lipger} & AV & 39.3$_{(-1.8\%)}$ & 34.2$_{(-6.3\%)}$ & 12.0$_{(-5.5\%)}$ & 13.4$_{(-5.6\%)}$ & 24.7$_{(-4.3\%)}$ \\
    GER w/ Auto-AVSR$^\dagger$ & AV & \textbf{18.9}$_{(-52.8\%)}$ & 39.0$_{(+6.8\%)}$ & 17.4$_{(+37.0\%)}$ & 18.1$_{(+27.5\%)}$ & 23.3$_{(-9.7\%)}$ \\
    \midrule
    RelPrompt (ours, w/o DualHyp) & A & 39.7$_{(-0.8\%)}$ & 33.6$_{(-7.9\%)}$ & 11.8$_{(-7.1\%)}$ & 13.1$_{(-7.7\%)}$ & 24.5$_{(-5.0\%)}$ \\
    RelPrompt (ours, w/o DualHyp) & V & - & - & - & - & 39.3$_{(+52.3\%)}$ \\
    \midrule
    \textbf{DualHyp} w/o LLM (ROVER)~\cite{fiscus1997post} & A\,+\,V & 29.4$_{(-26.5\%)}$ & 30.1$_{(-17.5\%)}$ & 13.8$_{(+8.7\%)}$ & 14.7$_{(+3.5\%)}$ & 22.0$_{(-14.7\%)}$ \\
    \textbf{DualHyp ({ours})} & A\,+\,V & 21.6$_{(-46.0\%)}$ & 17.9$_{(-51.0\%)}$ & 8.1$_{(-36.2\%)}$ & 9.3$_{(-34.5\%)}$ & 14.2$_{(-45.0\%)}$ \\
    \textbf{+\,RelPrompt ({ours})} & A\,+\,V & 20.4$_{(-49.0\%)}$ & \textbf{16.0}$_{(-56.2\%)}$ & \textbf{8.0}$_{(-37.0\%)}$ & \textbf{8.2}$_{(-42.3\%)}$ & \textbf{13.2}$_{(-48.8\%)}$ \\
    \bottomrule
    \addlinespace[2pt]
    \multicolumn{7}{c}{\textbf{(a) Audio: random noise [-10, 10]\,dB, Video: 50\% segment occluded with object}} \\
    \addlinespace[8pt]
    \toprule
    \textbf{Method} & \textbf{Input} & \textbf{Object} & \textbf{Hands} & \textbf{Pixelate} & \textbf{Blur} & \textbf{Overall} \\
    \midrule
    \color{gray}\textit{ASR oracle $o_{nb}$\,/\,$o_{cp}$} & \color{gray}A & \color{gray}- & \color{gray}- & \color{gray}- & \color{gray}- & \color{gray}10.9 / 6.6 \\
    \color{gray}{\textit{ASR\,+\,VSR oracle $o_{nb}$\,/\,$o_{cp}$}} & \color{gray}A\,+\,V & \color{gray}4.7 / 2.8 & \color{gray}4.5 / 2.6 & \color{gray}4.8 / 2.7 & \color{gray}4.1 / 2.4 & \color{gray}4.5 / 2.6 \\
    \midrule
    Whisper-large-v3 \cite{radford2023robust} & A & 26.7 & 26.7 & 26.7 & 26.7 & 26.7 \\
    BRAVEn-large \cite{haliassos2024braven} & V & 39.7$_{(+48.7\%)}$ & 35.1$_{(+31.5\%)}$ & 39.4$_{(+47.6\%)}$ & 31.7$_{(+18.7\%)}$ & 36.5$_{(+36.7\%)}$ \\
    \midrule
    GER~\cite{chen2023hyporadise} & A & - & - & - & - & 23.9$_{(-10.5\%)}$ \\
    RobustGER~\cite{hu2024robustger} & A & - & - & - & - & 24.9$_{(-6.7\%)}$ \\
    LipGER~\cite{ghosh2024lipger} & AV & 24.2$_{(-9.4\%)}$ & 24.3$_{(-9.0\%)}$ & 24.3$_{(-9.0\%)}$ & 24.1$_{(-9.7\%)}$ & 24.3$_{(-9.0\%)}$ \\
    GER w/ Auto-AVSR$^\dagger$ & AV & 29.5$_{(+10.5\%)}$ & 26.6$_{(-0.4\%)}$ & 29.1$_{(+9.0\%)}$ & 23.5$_{(-12.0\%)}$ & 27.2$_{(+1.9\%)}$ \\
    \midrule
    \textbf{DualHyp} w/o LLM (ROVER)~\cite{fiscus1997post} & A\,+\,V & 22.9$_{(-14.2\%)}$ & 21.9$_{(-18.0\%)}$ & 23.6$_{(-11.6\%)}$ & 20.7$_{(-22.5\%)}$ & 22.3$_{(-16.5\%)}$ \\
    \textbf{DualHyp ({ours})} & A\,+\,V & {12.0}$_{(-55.1\%)}$ & 11.8$_{(-55.7\%)}$ & 12.7$_{(-52.6\%)}$ & 11.1$_{(-58.4\%)}$ & 11.9$_{(-55.4\%)}$ \\
    \textbf{+\,RelPrompt ({ours})} & A\,+\,V & \textbf{11.9}$_{(-55.4\%)}$ & \textbf{11.0}$_{(-58.8\%)}$ & \textbf{11.9}$_{(-55.4\%)}$ & \textbf{10.2}$_{(-61.8\%)}$ & \textbf{11.3}$_{(-57.7\%)}$ \\
    \bottomrule
    \addlinespace[2pt]
    \multicolumn{7}{c}{\textbf{(b) Audio: speech noise 0\,dB, Video: random segment corrupted}}
  \end{tabular}
  }
  \vspace{-5pt}
  \caption{
  WER\% ($\downarrow$) results on the LRS2 test set under joint audio-visual corruption.
  (a) Performance across varying audio noise types, with a fixed visual corruption (50\% segment occluded by an object).
  (b) Performance across varying visual corruption types, with a fixed audio corruption (0\,dB speech noise).
  We also show the relative WER reduction in parentheses compared to the Whisper-large-v3 ASR baseline.
  All the ASR and VSR heads are Whisper-large-v3 and BRAVEn-large, respectively.
  $^\dagger$: We implement a GER model using hypotheses generated from an early-fusion approach, Auto-AVSR~\cite{ma2023auto}, which has been trained on LRS2 with babble noise.
  }
  \label{tab:main_lrs2}
\end{table*}

\section{Experiments}
\label{sec:experiments}

\subsection{Experimental Setup}
We conduct our experiments on the LRS2 AVSR benchmark~\cite{son2017lip} with the WER metric.
All models are trained and tested under diverse, synthetically corrupted audio-visual conditions, following the protocol of CAV2vec \cite{kim2025multi}.
Unless specified otherwise, our DualHyp framework is composed of a Whisper-large-v3 \cite{radford2023robust} ASR head, a {BRAVEn-large} \cite{haliassos2024braven} VSR head, and a {TinyLlama} \cite{zhang2024tinyllama} LLM, which we fine-tune using LoRA~\cite{hu2022lora}.
Full details of implementation, corruption protocol, and baseline methods are provided in Appendix~\ref{app:experimental_details}.
We also offer additional results on another AVSR benchmark, LRS3~\cite{afouras2018lrs3}, in Appendix~\ref{app:additional_results} to support solid performance.

\subsection{LRS2 Benchmark Results}

Table\,\ref{tab:main_lrs2} presents the benchmark results, where we isolate modality-specific robustness by either varying audio noise against fixed visual corruption (Table\,\ref{tab:main_lrs2}\textcolor{Red}{a}) or varying visual corruption against fixed audio noise (Table\,\ref{tab:main_lrs2}\textcolor{Red}{b}).
Our proposed \textbf{DualHyp\,+\,RelPrompt} achieves the lowest \textbf{overall WER of 13.2\%} under audio variability and \textbf{11.3\%} under visual variability, representing a relative improvement of 48.8\% and 57.7\% compared to the ASR baseline, Whisper-large-v3. 
This confirms our core hypothesis that LLMs can perform robust compositional reasoning when provided with separate ASR and VSR hypotheses.

In contrast, all baseline methods show clear limitations.
ASR-only models like GER \cite{chen2023hyporadise} or RobustGER \cite{hu2024robustger} are fundamentally capped by the input audio quality and struggle under low SNRs (also refer to \S\ref{subsec:snr_to_werr}).
Audio-visual approach like LipGER \cite{ghosh2024lipger} fails to improve over the standard GER framework, which shows that injecting video via additional adapter is insufficient for LLM to fully exploit the visual modality while harming its stability due to cross-modal gap~\cite{zhang2024llamaadapter, li2023prompting, gao2023llama}.
Similarly, GER w/ Auto-AVSR \cite{ma2023auto} exhibits strong but narrow performance, excelling only on the babble noise that the model's AVSR head is specifically trained on, failing to generalize to other conditions (see \S\ref{subsec:comparison_avsr} for further analysis).
Furthermore, while DualHyp without LLM reasoning (ROVER, \citet{fiscus1997post}) outperforms feature-level early fusion through text-level late fusion, it remains inferior to our full framework.
This indicates that simple voting mechanisms are insufficient to capture the semantic complementarity across multimodal hypotheses, underscoring the necessity of explicit LLM reasoning.

The success of our DualHyp with RelPrompt approach stems from two aspects: (1) a text-level late fusion strategy and (2) the ability to dynamically leverage the more reliable modality.
Our late fusion provides the LLM with rich, modality-specific evidence in a unified text format that is readily processed by the LLM.
The isolation of modalities also ensures that corruption in one stream does not contaminate the other. 
Then, RelPrompt dynamically leverages the more reliable stream, utilizing visual hypotheses when audio quality is low and falling back on audio hypotheses when the visual stream is degraded. An ablation study in Table\,\ref{tab:main_lrs2}\textcolor{Red}{a} confirms this, showing that RelPrompt yields only marginal gains in single-modality inference and intinsically relies on the dual-stream framework to facilitate cross-modal arbitration.
Notably, the overall success of this dual-stream approach is achieved even though our VSR model is substantially weaker than ASR, suggesting that our framework's potential is scalable as more powerful VSR models emerge.

\begin{table}[!t]
    \centering
    \small
    \addtolength{\tabcolsep}{1.5pt}
    \resizebox{\columnwidth}{!}{
    \begin{tabular}{lc|ccc}
        \toprule
        \textbf{Method} & \textbf{Input} & A$^c$V$^c$ & A$^c$V$^n$ & A$^n$V$^c$ \\
        \midrule
        Whisper-large-v3 & A & 3.8 & 3.8 & 25.8 \\
        BRAVEn-large & V & 26.9 & 36.5 & 26.9  \\
        GER & A & 2.6 & 2.6 & 24.6 \\
        \midrule
        \textbf{DualHyp} & A\,+\,V & \textbf{1.9} & 2.1 & 11.5 \\
        \textbf{+\,RelPrompt} & A\,+\,V & \textbf{1.9} & \textbf{2.0} & \textbf{9.9} \\
        \bottomrule
    \end{tabular}
    }
    \vspace{-5pt}
    \caption{Performance under different modality conditions on LRS2, with clean audio or video (X$^c$) and noisy audio or video (X$^n$), $\text{X}\in\{\text{A}, \text{C}\}$.}
    \label{tab:lrs2_clean}
\end{table}

\paragraph{Clean audio or video inputs.}
Even in the clean audio settings (Table~\ref{tab:lrs2_clean}), our DualHyp methods achieve the lowest WER, showing they effectively capitalize on the high-quality audio stream. In the noisy-audio/clean-video setting, while GER is severely hampered by corrupted audio (24.6\%), RelPrompt leverages clean visual hypotheses to dramatically improve to 9.9\%. The gap between DualHyp (11.5\%) and its reliability-guided version demonstrates that dynamically detecting clean signal (in this case video) and giving the LLM explicit hints about which to trust is effective.

\subsection{Larger LLMs}
\label{subsec:larger_llms}

We investigate the impact of LLM scale by evaluating our methods with three different models: TinyLlama~\cite{zhang2024tinyllama}, Phi-2~\cite{javaheripi2023phi2}, and Llama-3.2-3B~\cite{meta_llama3_2_2024}.
The results in Table~\ref{tab:larger_llm} show that the benefits of a larger LLM are most pronounced within our proposed framework.
For the GER baseline, scaling the LLM yields only marginal gains, indicating that the performance is limited by the quality of the single-stream input hypotheses.
In contrast, our models benefit more significantly from a larger LLM's capacity.
The effect is greatest for DualHyp + RelPrompt, which achieves the best overall WER of 12.3\% with Llama-3.2.
This suggests that by providing a richer and more comprehensive input, our framework creates a more sophisticated reasoning task that can effectively leverage the capabilities of LLMs.

\begin{table}[!t]
    \centering
    \small
    \addtolength{\tabcolsep}{-2pt}
    \resizebox{\columnwidth}{!}{
    \begin{tabular}{ll|cccc|c}
    \toprule
    \textbf{Method} & \textbf{LLM (Params.)} & \textbf{B} & \textbf{S} & \textbf{M} & \textbf{N} & \textbf{O} \\
    \midrule
    \multirow{3}{*}{GER} & TinyLlama (1.1B) & 39.3 & 34.4 & 11.5 & 13.2 & 24.6 \\
    & Phi-2 (2.7B) & 39.0 & 33.7 & 11.9 & 13.0 & 24.4 \\
    & Llama-3.2 (3.2B) & 38.9 & 34.1 & 11.6 & 12.9 & 24.4 \\
    \midrule
    \multirow{3}{*}{\textbf{DualHyp}} & TinyLlama (1.1B) & 21.6 & 17.9 & 8.1 & 9.3 & 14.2 \\
    & Phi-2 (2.7B) & 21.6 & 19.0 & 7.8 & 8.7 & 14.3 \\
    & Llama-3.2 (3.2B) & 20.4 & 16.0 & \textbf{7.2} & \textbf{8.1} & 12.9 \\    
    \midrule
    \multirow{3}{*}{\textbf{\thead[l]{DualHyp\\+\,RelPrompt}\!\!}} & TinyLlama (1.1B) & 20.4 & 16.0 & 8.0 & 8.2 & 13.2 \\
    & Phi-2 (2.7B) & 21.1 & 18.2 & 8.0 & 8.5 & 14.0 \\
    & Llama-3.2 (3.2B) & \textbf{19.6} & \textbf{14.1} & 7.4 & 8.2 & \textbf{12.3} \\
    \bottomrule
    \end{tabular}
    }
    \vspace{-5pt}
    \caption{WER (\%) comparison using different LLMs on the LRS2 benchmark.
    The corruption strategy follows Table\,\ref{tab:main_lrs2}\textcolor{Red}{a}, where \textbf{B}, \textbf{S}, \textbf{M}, and \textbf{N} represent each noise type with the overall result (\textbf{O}).}
    \label{tab:larger_llm}
\end{table}

\begin{table}[!t]
    \centering
    \small
    \addtolength{\tabcolsep}{2.0pt}
    \resizebox{\columnwidth}{!}{
    \begin{tabular}{l|cccc|c}
    \toprule
    \textbf{Method} & \textbf{Es} & \textbf{Fr} & \textbf{It} & \textbf{Pt} & \textbf{Avg} \\
    \midrule
    Whisper-large-v3 & 49.6 & \textbf{46.8} & 52.3 & 52.7 & 50.4 \\
    mAV-HuBERT & 70.5 & 81.7 & 73.7 & 74.1 & 75.0 \\
    GER & 50.6 & 47.8 & 58.5 & 52.3 & 52.3 \\
    \textbf{DualHyp} & \textbf{47.3} & 47.9 & \textbf{47.2} & \textbf{49.0} & \textbf{47.9} \\
    \bottomrule
    \end{tabular}
    }
    \vspace{-5pt}
    \caption{WER (\%) comparison with multilingual babble noise (SNR\,=\,0\,dB) on the MuAViC dataset.}
    \label{tab:muavic}
\end{table}

\begin{figure*}
    \centering
    \includegraphics[width=\linewidth]{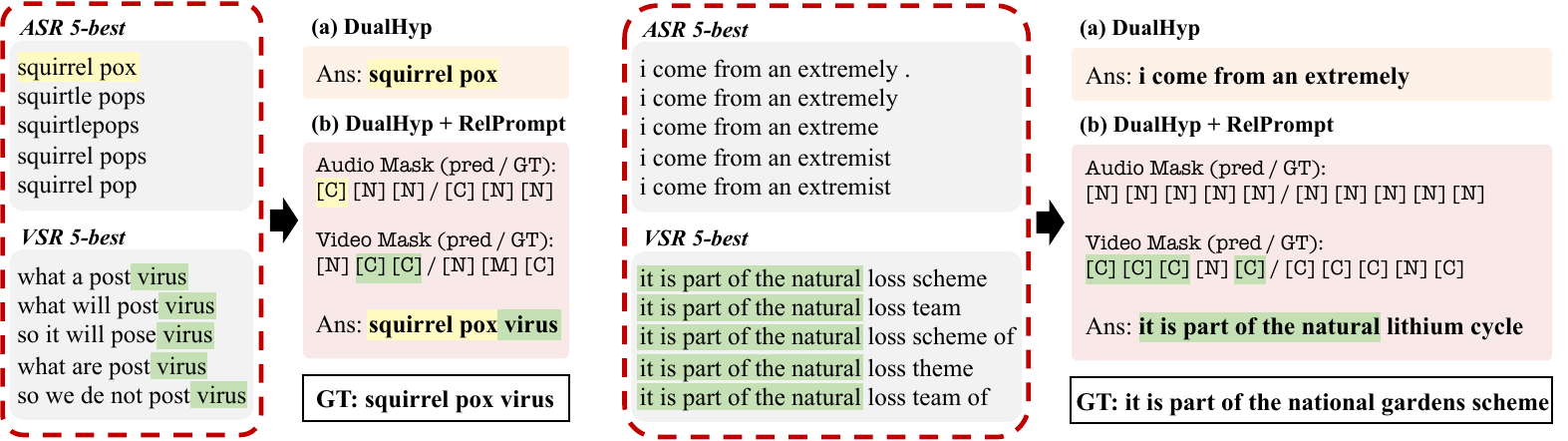}
    \caption{Qualitative analysis comparing RelPrompt to the DualHyp baseline.
    RelPrompt uses reliability tokens (\ie, masks) to explicitly inform the input signal quality, correctly guiding the use of ASR and VSR hypotheses.
    }
    \label{fig:qual_analysis}
\end{figure*}

\subsection{Multilingual AVSR}

To evaluate our framework in a multilingual context, we conduct experiments on the MuAViC dataset~\cite{anwar2023muavic} with adding multilingual babble noise at SNR 0\,dB~\cite{kim2025mohave}.
While the Whisper ASR head remains the same as in prior experiments, a VSR head is fine-tuned from mAV-HuBERT~\cite{kim2024efficient} for each language, due to the absence of strong multilingual VSR system.
Llama-3.2-3B is employed for the multilingual reasoning.
In Table~\ref{tab:muavic}, our framework outperforms both Whisper and GER in three of the four languages. %
However, this performance gain can be limited when VSR performance is severely degraded, as observed in the French case.
We thus anticipate that the performance gains of our methodology will become even more significant as more powerful multilingual VSR models emerge.

\section{Analysis}
\label{sec:analysis}

\subsection{Qualitative Analysis}

Our qualitative analysis in Figure~\ref{fig:qual_analysis} illustrates how RelPrompt corrects failures of the baseline DualHyp framework by providing explicit reliability signals.
(\textit{Left}): RelPrompt uses clean video tokens \texttt{[C]} as a cue to trust the last part of the VSR hypotheses, allowing it to recover the word (\texttt{virus}) which the baseline has missed.
(\textit{Right}): The ASR system is presented with fluent but entirely incorrect hypotheses.
By referencing the consistently noisy audio tokens \texttt{[N]}, the LLM correctly identifies the ASR stream as unreliable and pivots to the more accurate VSR candidates.
In contrast, without the RelPrompt mechanism, the model lacks any modality-level grounding and produces a completely incorrect output.
These cases demonstrate that by providing explicit reliability tokens, RelPrompt empowers the LLM to act as an intelligent controller, grounding its compositional reasoning in the predicted quality of the source signals.

\begin{table}[!t]
    \centering
    \small
    \addtolength{\tabcolsep}{2pt}
    \resizebox{\columnwidth}{!}{
    \begin{tabular}{c|cccc|c}
        \toprule
        \textbf{SNR} & \textbf{Acc.} & \!\!\textbf{Precision}\!\! & \textbf{Recall} & \textbf{F1} & \textbf{WER} \\
        \midrule
        $-$10\,dB & 84.7 & 95.3 & 87.8 & 91.4 & 25.8 \\
        $-$5\,dB & 83.9 & 95.0 & 87.1 & 90.9 & 17.8 \\
        0\,dB & 82.2 & 94.4 & 85.5 & 89.7 & 7.2 \\
        5\,dB & 79.6 & 93.0 & 82.8 & 87.6 & 3.4 \\
        10\,dB & 76.2 & 90.9 & 78.2 & 84.1 & 2.5 \\
        \bottomrule
    \end{tabular}
    }
    \vspace{-5pt}
    \caption{Performance (\%) of the reliability mask predictors with randomly corrupted audio and video segments. %
    The metrics evaluate the classification of segments as \texttt{noisy}, which includes the \texttt{mixed} category.}
    \label{tab:mask_acc}
\end{table}

\subsection{Reliability Mask Prediction}

In Table~\ref{tab:mask_acc}, our evaluation of the reliability predictors reveals two key strengths.
First, the predictor shows consistently high precision (>90\%), which ensures that its \texttt{noisy} flags are highly trustworthy and prevents the main model from incorrectly discarding clean data.
Second, the recall naturally decreases as the SNR increases.
This is a desirable behavior, as the predictor conservatively labels mildly corrupted audio segments as \texttt{clean}, allowing the model to continue exploiting the useful signal.

\begin{table}[!t]
    \centering
    \small
    \addtolength{\tabcolsep}{-1pt}
    \resizebox{\columnwidth}{!}{
    \begin{tabular}{lcc|cccc|c}
        \toprule
        \textbf{Method} & \!\textbf{Input}\! & \!\!\textbf{\# hyps}\!\! & \textbf{B} & \textbf{S} & \textbf{M} & \textbf{N} & \textbf{O} \\
        \midrule
        \multirow{3}{*}{GER} & A & 5 & 39.3 & 34.4 & 11.5 & 13.2 & 24.6 \\
         & AV & 5 & 18.9 & 39.0 & 17.4 & 18.1 & 23.3 \\
         & AV & 10 & 18.2 & 38.1 & 16.7 & 17.6 & 22.6 \\
        \midrule
        \multirow{2}{*}{\textbf{DualHyp}} & \!\!A\,+\,AV\!\! & 10 & {17.1} & 26.7 & \textbf{7.2} & {8.4} & 14.8 \\
        & A\,+\,V & 10 & 21.6 & {17.9} & 8.1 & 9.3 & {14.2} \\
        \midrule
        \textbf{DualHyp} & \!\!A\,+\,AV\!\! & 10 & \textbf{15.4} & 25.9 & 7.3 & 8.8 & 14.3 \\
        \textbf{+\,RelPrompt} & A\,+\,V & 10 & 20.4 & \textbf{16.0} & 8.0 & \textbf{8.2} & \textbf{13.2} \\
        \bottomrule
    \end{tabular}
    }
    \vspace{-5pt}
    \caption{   
    WER (\%) comparison of different hypotheses from single-stream (GER) and dual-stream (DualHyp) generation heads.
    Note that the AVSR head is trained on LRS2 with babble noise~\cite{ma2023auto}, unlike the ASR and VSR heads.
    }
    \label{tab:comparison_avsr}
\end{table}

\subsection{Comparison with an AVSR Head}
\label{subsec:comparison_avsr}

Our analysis in Table~\ref{tab:comparison_avsr} highlights two key findings regarding hypothesis generation.
First, modality diversity of hypotheses is more crucial than sheer quantity.
Simply increasing the number of hypotheses for the single-stream GER (5\,$\rightarrow$\,10 AV hypotheses) yields only a marginal gain for overall performance (23.3\%\,$\rightarrow$\,22.6\%), compared to DualHyp using 5-best hypotheses from each distinct modality (23.3\%\,$\rightarrow$\,14.2\%).

Second, while AVSR hypotheses might seem viable alternatives to VSR, they remain overly dependent on the audio modality.
This is particularly evident under the speech noise condition, where the visual stream is crucial for disambiguating target utterance from interfering speech.
In this scenario, DualHyp (A\,+\,AV) struggles (26.7\% WER), as the early fusion of AVSR embeddings makes visual information rely on the corrupted audio.
Instead, DualHyp (A\,+\,V) leverages the audio-independent VSR stream to achieve 17.9\%, demonstrating the superiority of using disentangled hypotheses.
These findings are further supported by our LRS3 experiments (Table~\ref{tab:comparison_avsr_lrs3}).

\begin{figure}[!t]
    \centering
    \includegraphics[width=\linewidth]{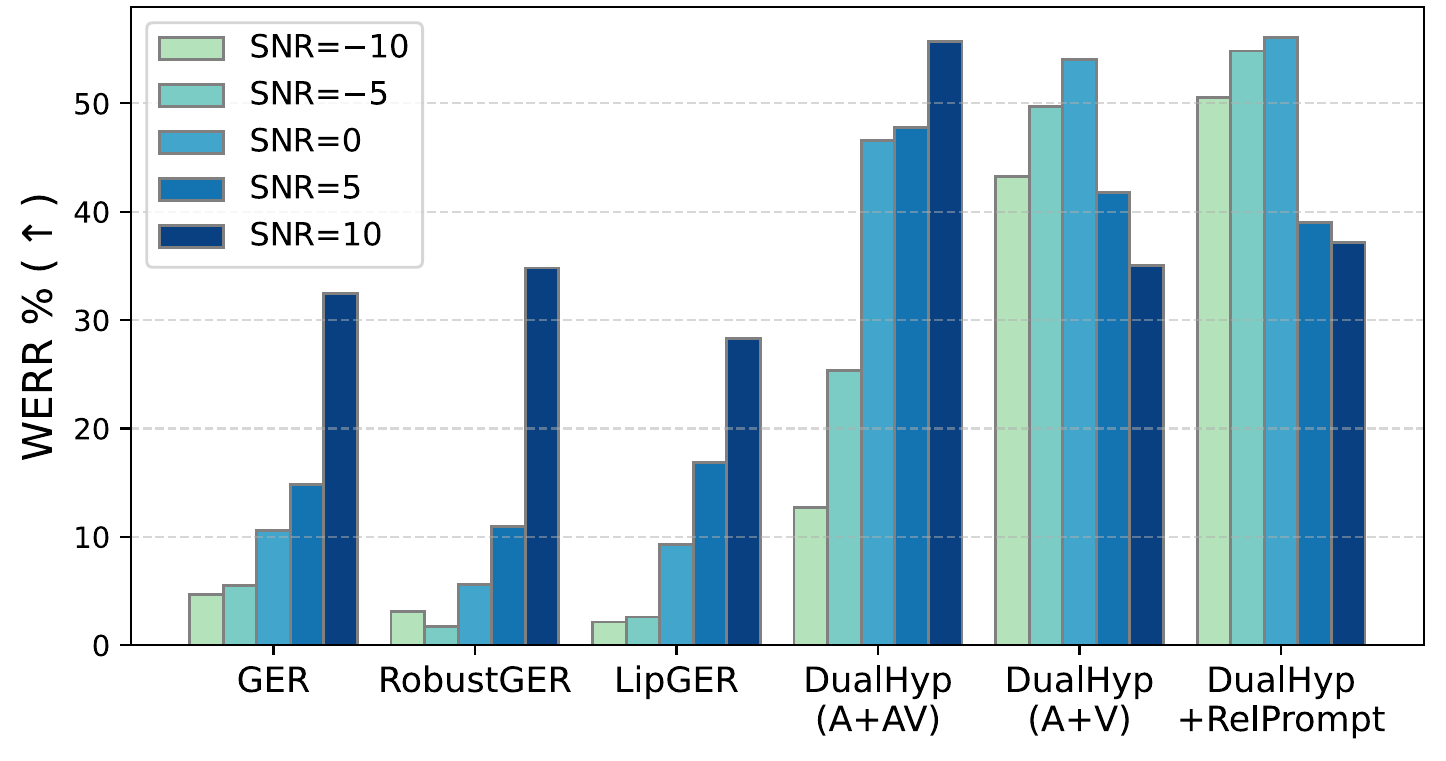}
    \vspace{-23pt}
    \caption{WERR at different audio SNRs, under speech noise. Higher WERR indicates greater improvement over the Whisper ASR baseline. 
    }
    \label{fig:snr_analysis}
\end{figure}

\subsection{SNR-wise WER Improvement}
\label{subsec:snr_to_werr}

Figure~\ref{fig:snr_analysis} reveals opposing trends in WER reduction (WERR, \citet{liu2025avger}) between single-stream and dual-stream methods.
For single-stream methods, WERR increases with better audio quality, as their effectiveness is limited to refining an already decent ASR output.
In contrast, our dual-stream framework maintains a high WERR even at very low SNRs by leveraging VSR hypotheses. %
Furthermore, the addition of RelPrompt consistently boosts performance, with the most significant gains observed in low-SNR scenarios.
This confirms that by effectively utilizing the reliability information about corruption provided by RelPrompt, our framework can substantially reduce errors precisely when the audio is most challenging.

\section{Conclusion}
In this study, we introduced DualHyp, a novel GER framework for AVSR that deliberately delays modality fusion to the language space, where an LLM performs compositional reasoning on independent hypotheses from ASR and VSR models. We further enhanced this with RelPrompt, a noise-aware guidance mechanism that guides the LLM with explicit, time-aligned reliability signals for each modality. The experiments showed that our new framework significantly outperforms single-stream GER approaches, highlighting a flexible paradigm that leverages modular integration.

\section*{Limitations}
While our framework demonstrates significant robustness and scalability in AVSR, it still holds two primary limitations that are common to most GER systems.

First, the performance of our framework is fundamentally dependent on the quality of its consisting components, especially the upstream SR heads. If the initial hypotheses from the SR head are of poor quality, as seen in our results of the MuAViC French case, the LLM's ability to perform corrections is limited. This dependency currently restricts the framework's applicability beyond English, because there is no publicly available, high-quality multilingual VSR model, making adaptation to the low-resource speech recognition and translation challenging.

Second, multiple modules in our structure introduces computational latency, posing a challenge for real-time applications. Although the ASR and VSR streams can be processed in parallel, the final LLM correction step is sequential, creating an unavoidable bottleneck. While modern efficiency techniques like flash attention can mitigate this to an extent, the approach remains inherently slower than a single end-to-end model, making deployment on resource-constrained edge devices a significant hurdle.

\section*{Acknowledgments}
This work was supported by Institute of Information \& communications Technology Planning \& Evaluation (IITP) grant funded by the Korea government(MSIT) [No.2022-0-00641, XVoice: Multi-Modal Voice Meta Learning] and by the InnoCORE program of the Ministry of Science and ICT (N10250156).

\bibliography{custom}

\appendix

\clearpage

{
\Large
\noindent
\textbf{Appendix}
}

\section{Release of DualHyp Dataset}
To facilitate future research within our DualHyp framework, we publicly release the hypotheses dataset\footnote{\url{https://github.com/sungnyun/dualhyp}}.
The primary motivation of this dataset construction is to decouple the computationally expensive hypothesis generation step from the LLM fine-tuning process. By providing pre-generated ASR and VSR hypotheses, this dataset will allow researchers to focus directly on developing novel language-space fusion and correction strategies, significantly lowering the barrier to entry.

Our DualHyp hypotheses are mainly created from LRS2~\citep{son2017lip} and LRS3~\citep{afouras2018lrs3} datasets. LRS2 is a benchmark of British English speech from BBC that covers diverse speakers and topics, while LRS3 consists of spoken utterances from TED and TEDx recordings.
For LRS2, our hypotheses dataset covers the standard splits (45,830 training, 1,082 validation, and 1,243 test utterances). For high-resource training (dealt in Appendix~\ref{subsec:high-reource_training}), we use additional 95,642 utterances, and for the LRS3 experiments (dealt in Appendix~\ref{app:lrs3_result}), we use 30,775 training utterances. 

For both the ASR head\footnote{\url{https://huggingface.co/openai/whisper-large-v3}} and VSR head\footnote{\url{https://github.com/ahaliassos/raven}}, we generate hypotheses using a beam size of 50. We select the 5-best unique hypotheses from each stream. If fewer than five unique hypotheses are generated, we randomly sample from the existing ones to reach the size 5. This results in a total of 10 hypotheses (5 from ASR, 5 from VSR) that are fed into the LLM for error correction.

Each entry also includes the ground-truth transcription and metadata detailing the specific audio or visual corruption applied (see Appendix\,\ref{sec:corruption_protocol}). Because different corruption types result in different output hypotheses, we separately save the dataset for each corruption condition. For training, a complete dataset is formed by merging these individual sets and randomly sampling hypotheses.

\section{Experimental Details}
\label{app:experimental_details}

\subsection{Implementation of DualHyp}

We fine-tune the LLM using LoRA~\cite{hu2022lora} with a rank of $r=16$. The number of trainable parameters is 4.5M for TinyLlama\footnote{\url{https://huggingface.co/TinyLlama/TinyLlama-1.1B-Chat-v1.0}}, 23.6M for Phi-2\footnote{\url{https://huggingface.co/microsoft/phi-2}}, and 24.3M for Llama-3.2\footnote{\url{https://huggingface.co/meta-llama/Llama-3.2-3B}}. For TinyLlama, we apply LoRA to the attention layers (key, value, query, and projection) only. For the larger Phi-2 and Llama-3.2 models, we apply LoRA to both the attention module and the feed-forward network (FFN) layers to ensure better convergence. All models are trained for 5 epochs with a batch size of 32 and a learning rate of 1e-4.

For the implementation of RelPrompt, our reliability predictors are designed lightweight, with only 1.1M parameters each. The architecture consists of two 1D-convolutional layers followed by average pooling and a final linear classifier to match the segment size. For training these predictors, we create ground-truth labels for each 0.4-second segment based on its constituent frames: a segment is labeled \texttt{[C]} (Clean) if less than 10\% of its frames are corrupted, \texttt{[N]} (Noisy) if more than 60\% of its frames are corrupted, and \texttt{[M]} (Mixed) otherwise.

The full DualHyp\,+\,RelPrompt model is trained for 8 hours on a single NVIDIA A6000 GPU, using a learning rate of 2e-4 for the main LLM (with LoRA) and 1e-4 for the reliability predictors. For all data pre-processing and evaluation, we use the publicly available packages following the LipGER codebase\footnote{\url{https://github.com/Sreyan88/LipGER}}.

\subsection{Corruption Protocol}
\label{sec:corruption_protocol}
In our study, all models are trained and evaluated under challenging noisy conditions to assess their robustness in real-world scenarios.
To ensure a robust evaluation, we introduce a diverse set of synthetic corruptions into the LRS2 dataset, following the protocol established by \citet{kim2025multi}\footnote{\url{https://github.com/sungnyun/cav2vec}}.
\begin{itemize}[leftmargin=*, itemsep=0pt]
    \item Audio corruptions: We augment the audio streams with four types of noise, similar to \citet{shi2022robust}. We use speech noise from the LRS3 dataset~\cite{afouras2018lrs3} and babble, music, and natural sounds from the MUSAN corpus~\cite{snyder2015musan}.
    \item Visual corruptions: We apply four common visual degradation types: object occlusion~\cite{voo2022delving}, hands occlusion, pixelation, and blur~\cite{kim2025multi}.
\end{itemize}

During training, we randomly apply one of these corruption types to each sample. The duration of the applied corruption is also randomized, with its portion sampled from a Beta distribution ($\alpha,\beta=2.0$) to simulate varying levels of interference. 

For evaluation, we apply background noise to the entire audio sample and corrupt partial video segments to better reflect real-world scenarios. 
For Table~\ref{tab:main_lrs2}\textcolor{red}{a}, audio noise is applied to the entire time duration with SNR randomly sampled from [-10, 10]\,dB, while half of the video segments is occluded with object. 
For Table~\ref{tab:main_lrs2}\textcolor{red}{b}, 0\,dB SNR of speech noise is augmented to the whole audio, while video corruption length is sampled from Beta distribution.

To assess overall performance, we report the average WER from a single comprehensive evaluation run. This run covers all test samples and incorporates a diverse range of noisy conditions to ensure the statistical credibility of our methods.

\subsection{Baseline Methods}
For a fair comparison, we train all baseline methods from scratch on the same set of corrupted audio-visual data. We have found that this diverse noise training significantly boosts the performance of all GER-based methods, which establishes a strong set of baselines for our evaluation. Our primary baselines are:
\begin{itemize}[leftmargin=*, itemsep=0pt]
    \item GER \cite{chen2023hyporadise}: The foundational LLM-based error correction framework that operates on \textit{N}-best hypotheses from an ASR model.
    \item RobustGER \cite{hu2024robustger}: An extension of GER designed to improve robustness against noisy audio conditions.
    \item LipGER \cite{ghosh2024lipger}: An audio-visual GER method that incorporates visual features via an adapter but still relies on a single stream of ASR hypotheses for correction.
    \item GER w/ Auto-AVSR \cite{ma2023auto}: A strong baseline we implement by feeding the \textit{N}-best hypotheses from the early-fusion Auto-AVSR model into a standard GER framework.
    \item ROVER \cite{fiscus1997post}: A post-processing ensemble framework that constructs a word transition network through the iterative alignment of multiple hypotheses.
\end{itemize}

We note that training these single-stream GER baselines presents a significant stability issue when using highly corrupted data. As detailed in our analysis (\S\ref{subsec:snr_to_werr}), the performance of these models is capped by the quality of the initial ASR hypotheses. During training, low-SNR audio produces poor learning signal, which causes the model to learn to over-correct already accurate transcriptions while failing to fix genuinely erroneous ones. We have observed their performance degradation (up to +5\% WER) when trained with the same data as DualHyp. To ensure stable convergence for our baseline comparisons, we therefore construct their training dataset exclusively from audio samples with SNR\,$\ge$\,0\,dB, just as \citet{hu2024robustger, ghosh2024lipger} have constructed their training datasets.

Our ROVER implementation conducts iterative alignment of hypotheses based on the 1-best hypothesis to construct a word transition network.
We employ majority voting for each column with \texttt{<NULL>} tokens to prevent excessive insertion errors.
Tie-breaking is resolved based on the rank order of the input hypothesis lists.
No confidence scores or external LM weights are employed, as our setup relies solely on text hypotheses.

\section{DualHyp Analysis}
\label{app: qualitative_examples}

\begin{table*}[!t]
\centering
\small
\setlength{\fboxsep}{1pt} %
\resizebox{\textwidth}{!}{
\begin{tabularx}{\linewidth}{l|X|c}
\toprule
\textbf{Method} & \textbf{Utterance} & \textbf{WER (\%)} \\
\midrule
\midrule
\multicolumn{3}{c}{\textit{\textbf{Type 1: Multimodal Fragment Composition}}} \\
\midrule
\multirow{5}{*}{ASR 5-best} &
    \highlight{yellow!30}{which upset some}one in the next day & 71.4  \\
    & \highlight{yellow!30}{which} i am saying some of you may or may not understand & 128.6 \\
    & \highlight{yellow!30}{which upset some}one who knew not what to do & 100.0 \\
    & \highlight{yellow!30}{which} i am saying is a lot easier than it is today & 157.1 \\
    & \highlight{yellow!30}{which upset some}one you know what i mean & 85.7 \\
    \midrule
\multirow{5}{*}{VSR 5-best} &
    we jumped at \highlight{orange!20}{some of our female}\highlight{Green!10}{ residents} & 42.9 \\
    & we jumped at \highlight{orange!20}{some of our female} races & 57.1 \\
    & we jumps at \highlight{orange!20}{some of our female}\highlight{Green!10}{ residents} & 42.9 \\
    & we jump set \highlight{orange!20}{some of our female}\highlight{Green!10}{ residents} & 42.9 \\
    & we jumped at \highlight{orange!20}{some of our female} reasons & 57.1 \\
\midrule
\textbf{DualHyp output} &
\textbf{\highlight{yellow!30}{which upset some}\highlight{orange!20}{ of our female}\highlight{Green!10}{ residents}} & \textbf{0.0} \\
\midrule
Ground-truth & which upset some of our female residents & -- \\
\midrule
\midrule
\multirow{5}{*}{ASR 5-best} &
    \highlight{yellow!30}{so} what are the dangers of\highlight{purple!15}{ relying on}\highlight{blue!10}{ this information} & 62.5 \\
    & \highlight{yellow!30}{so} what are the dangers of\highlight{purple!15}{ relying on} disinformation & 87.5 \\
    & \highlight{yellow!30}{so} my other thing is\highlight{Green!10}{ just}\highlight{purple!15}{ relying on}\highlight{blue!10}{ this information} & 50.0 \\
    & \highlight{yellow!30}{so} what are the dangers\highlight{purple!15}{ relying on}\highlight{blue!10}{ this information} & 50.0 \\
    & \highlight{yellow!30}{so} my other thing is\highlight{purple!15}{ relying on}\highlight{blue!10}{ this information} & 50.0 \\
\midrule
\multirow{5}{*}{VSR 5-best} &
    \highlight{yellow!30}{so}\highlight{orange!20}{ rather than}\highlight{Green!10}{ just} regarding all\highlight{blue!10}{ this information} & 25.0 \\
    & \highlight{yellow!30}{so}\highlight{orange!20}{ rather than} regarding all\highlight{blue!10}{ this information} & 37.5 \\
    & \highlight{yellow!30}{so}\highlight{orange!20}{ rather than} to regard all\highlight{blue!10}{ this information} & 37.5 \\
    & \highlight{yellow!30}{so}\highlight{orange!20}{ rather than}\highlight{Green!10}{ just} regarding\highlight{blue!10}{ this information} & 25.0 \\
    & \highlight{yellow!30}{so}\highlight{orange!20}{ rather than} argue\highlight{blue!10}{ this information} & 37.5 \\
\midrule
\textbf{DualHyp output} &
\textbf{\highlight{yellow!30}{so}\highlight{orange!20}{ rather than}\highlight{Green!10}{ just}\highlight{purple!15}{ relying on}\highlight{blue!10}{ this information}} & \textbf{0.0} \\
\midrule
Ground-truth & so rather than just relying on this information & -- \\
\midrule
\midrule
\multicolumn{3}{c}{\textit{\textbf{Type 2: Dominant Modality Refinement}}} \\
\midrule
\multirow{5}{*}{ASR 5-best} &
    \highlight{yellow!30}{the armed forces} were & 33.3 \\
    & \highlight{yellow!30}{the armed forces} go & 33.3 \\
    & and\highlight{yellow!30}{ the armed forces} were & 66.7 \\
    & and\highlight{yellow!30}{ the armed forces} go & 66.7 \\
    & in\highlight{yellow!30}{ the armed forces} but & 66.7 \\
\midrule
\multirow{5}{*}{VSR 5-best} &
    i feel disco & 100.0 \\
    & helpful disco & 100.0 \\
    & i fell this & 100.0 \\
    & time for disco & 100.0 \\
    & helpful to this & 100.0 \\
\midrule
\textbf{DualHyp output} & \textbf{\highlight{yellow!30}{the armed forces}} & \textbf{0.0} \\
\midrule
Ground-truth & the armed forces & -- \\
\midrule
\midrule
\multirow{5}{*}{ASR 5-best} &
    is \highlight{yellow!30}{already}\highlight{orange!20}{ in}\highlight{Green!10}{ the}\highlight{purple!15}{ states} & 25.0 \\
    & \highlight{orange!20}{in}\highlight{Green!10}{ the} united\highlight{purple!15}{ states} & 50.0 \\
    & of\highlight{Green!10}{ the} united\highlight{purple!15}{ states} & 75.0 \\
    & from the rest of\highlight{Green!10}{ the} united\highlight{purple!15}{ states} & 100.0 \\
    & \highlight{orange!20}{in} one of\highlight{Green!10}{ the} other\highlight{purple!15}{ states} & 75.0 \\
\midrule
\multirow{5}{*}{VSR 5-best} &
    \highlight{yellow!30}{already} understand & 75.0 \\
    & \highlight{yellow!30}{already} understanding & 75.0 \\
    & \highlight{yellow!30}{already} understands & 75.0 \\
    & \highlight{yellow!30}{already} understanding that & 100.0 \\
    & we are writing these things & 125.0 \\
\midrule
\textbf{DualHyp output} &
\textbf{\highlight{yellow!30}{already}\highlight{orange!20}{ in}\highlight{Green!10}{ the}\highlight{purple!15}{ states}} & \textbf{0.0} \\
\midrule
Ground-truth & already in the states & -- \\
\bottomrule
\end{tabularx}
}
\caption{Successful examples of the GER process using DualHyp. Highlights illustrate how the final output is assembled from partial information scattered across the ASR (audio) and VSR (video) 5-best hypothesis lists. These cases illustrate two primary successful patterns: multi-modal fragment composition and dominant modality refinement.}
\label{tab:successful_cases}
\end{table*}
\begin{table*}[!t]
\centering
\small
\setlength{\fboxsep}{1pt} %
\resizebox{\textwidth}{!}{
\begin{tabularx}{\linewidth}{l|X|c}
\toprule
\textbf{Method} & \textbf{Utterance} & \textbf{WER (\%)} \\
\midrule
\midrule
\multicolumn{3}{c}{\textit{\textbf{Type 1: Over-reliance on Plausible but Inaccurate Hypotheses}}} \\
\midrule
\multirow{5}{*}{ASR 5-best} &
    what could be\highlight{Green!10}{ in the world} & 75.0 \\
    & to be\highlight{Green!10}{ in the world} & 50.0 \\
    & i can not believe\highlight{Green!10}{ the world} & 100.0 \\
    & it should be\highlight{Green!10}{ in the world} & 75.0 \\
    & good to be\highlight{Green!10}{ in the world} & 75.0 \\
\midrule
\multirow{5}{*}{VSR 5-best} &
    \highlight{red!10}{what we do}\highlight{red!30}{ is} & 100.0 \\
    & \highlight{red!10}{when we do} this & 125.0 \\
    & \highlight{red!10}{what we do}\highlight{red!20}{ here}\highlight{red!30}{ is} & 100.0 \\
    & \highlight{red!10}{what we do} with this & 125.0 \\
    & what we are doing\highlight{red!30}{ is} & 100.0 \\
\midrule
\textbf{DualHyp output} &
\textbf{\highlight{red!10}{what we do}\highlight{red!20}{ here}\highlight{red!30}{ is}} & \textbf{125.0} \\
\midrule
Ground-truth & \textbf{probably\highlight{Green!10}{ in the world}} & -- \\
\midrule
\midrule
\multirow{5}{*}{ASR 5-best} &
    that is what i am talking about & 100.0 \\
    & i\highlight{Green!20}{ can not} believe it & 100.0 \\
    & \highlight{Green!10}{i just}\highlight{Green!20}{ can not} believe it & 42.9 \\
    & \highlight{Green!10}{i} am \highlight{Green!10}{just} going to come with you & 85.7 \\
    & that is what i am saying & 114.3 \\
\midrule
\multirow{5}{*}{VSR 5-best} &
    \highlight{Green!10}{i just}\highlight{red!10}{ asked them}\highlight{Green!30}{ to win} & 42.9 \\
    & \highlight{Green!10}{i just}\highlight{red!10}{ asked} him \highlight{Green!30}{ to win} & 42.9 \\
    & \highlight{Green!10}{i just}\highlight{red!10}{ asked them}\highlight{red!20}{ to wait} & 57.1 \\
    & \highlight{Green!10}{i just}\highlight{red!10}{ asked} him\highlight{red!20}{ to wait} & 57.1 \\
    & \highlight{Green!10}{i just}\highlight{red!10}{ ask them}\highlight{Green!30}{ to win} & 42.9 \\
\midrule
\textbf{DualHyp output} &
\textbf{\highlight{Green!10}{i just}\highlight{red!10}{ asked them}\highlight{red!20}{ to wait}} & \textbf{57.1} \\
\midrule
Ground-truth & \textbf{\highlight{Green!10}{i just}\highlight{Green!20}{ can not} seem\highlight{Green!30}{ to win}} & -- \\
\midrule
\midrule
\multicolumn{3}{c}{\textit{\textbf{Type 2: Hallucination and Semantic Association Errors}}} \\
\midrule
\multirow{5}{*}{ASR 5-best} &
    no no no & 100.0 \\
    & no no very good & 166.7 \\
    & love love love & 100.0 \\
    & no no & 100.0 \\
    & i love \highlight{Green!10}{november} & 66.7 \\
\midrule
\multirow{5}{*}{VSR 5-best} &
    it goes \highlight{Green!10}{november} & 66.7 \\
    & \highlight{Green!10}{november} & 66.7 \\
    & it is on \highlight{Green!10}{november} & 66.7 \\
    & it is not \highlight{Green!10}{november} & 66.7 \\
    & it is called \highlight{Green!10}{november} & 66.7 \\
\midrule
\textbf{DualHyp output} &
\textbf{\highlight{Green!10}{november}\highlight{red!10}{ and december}} & \textbf{100.0} \\
\midrule
Ground-truth & \textbf{end of \highlight{Green!10}{november}} & -- \\
\midrule
\midrule
\multirow{5}{*}{ASR 5-best} &
    \highlight{Green!10}{this is the best} bathroom downtown & 50.0 \\
    & \highlight{Green!10}{this is the best} bathroom town\highlight{Green!10}{ in town} & 25.0 \\
    & \highlight{Green!10}{this is the best} bathroom\highlight{Green!10}{ in town} & 25.0 \\
    & \highlight{Green!10}{this is the best} bath\highlight{Green!10}{ hotel in town} & 12.5 \\
    & \highlight{Green!10}{this is the best} basketball town\highlight{Green!10}{ in town} & 50.0 \\
\midrule
\multirow{5}{*}{VSR 5-best} &
    \highlight{Green!10}{this is the best} bad\highlight{Green!10}{ hotel in town} & 12.5 \\
    & \highlight{Green!10}{this is the best} band\highlight{Green!10}{ hotel in town} & 12.5 \\
    & \highlight{Green!10}{this is the best bat hotel in town} & 0.0 \\
    & \highlight{Green!10}{this is the best} pat\highlight{Green!10}{ hotel in town} & 12.5 \\
    & \highlight{Green!10}{this is the best} baton\highlight{Green!10}{ hotel in town} & 12.5 \\
\midrule
\textbf{DualHyp output} &
\textbf{\highlight{Green!10}{this is the best}\highlight{red!10}{ bistro}\highlight{Green!10}{ in town}} & \textbf{25.0} \\
\midrule
Ground-truth & \textbf{\highlight{Green!10}{this is the best bat hotel in town}} & -- \\
\bottomrule
\end{tabularx}
}
\caption{Failure examples of the GER process using DualHyp. Green highlights illustrate the correct words from ground-truth, whereas red highlights illustrate wrong words from inference. These cases illustrate two primary error patterns: over-reliance on plausible but inaccurate hypotheses and hallucination based on semantic association.}
\label{tab:failure_cases}
\end{table*}

\begin{table*}[!t]
\centering
\small
\setlength{\fboxsep}{1pt} %
\resizebox{\textwidth}{!}{
\begin{tabularx}{\linewidth}{l|X|c}
\toprule
\textbf{Method} & \textbf{Utterance} & \textbf{WER (\%)} \\
\midrule
\midrule
\multirow{5}{*}{ASR 5-best} &
    to your baby of this year when she asked & 100.0  \\
    & to your baby of this year when she asks & 100.0 \\
    & \highlight{yellow!30}{there was} your baby of this year when she asked & 100.0 \\
    & there is your baby of this year when she asked & 88.9 \\
    & to your baby of this year when she asked & 100.0 \\
    \midrule
\multirow{5}{*}{VSR 5-best} &
    \highlight{yellow!30}{there was}\highlight{orange!20}{ no} air \highlight{Green!10}{so there was no sound} & 22.2 \\
    & \highlight{yellow!30}{there was}\highlight{orange!20}{ no} hit \highlight{Green!10}{so there was no sound} & 33.3 \\
    & \highlight{yellow!30}{there was}\highlight{orange!20}{ no} heat \highlight{Green!10}{so there was no sound} & 33.3 \\
    & \highlight{yellow!30}{there was}\highlight{orange!20}{ no} heart there was no sound & 44.4 \\
    & \highlight{yellow!30}{there was}\highlight{orange!20}{ no} it \highlight{Green!10}{so there was no sound} & 33.3 \\
\midrule
\textbf{DualHyp output} & \textbf{\highlight{yellow!30}{there was} your baby of this year when she asked} & \textbf{100.0} \\
\midrule
\multirow{5}{*}{\textbf{RelPrompt output}} & \texttt{Audio Mask (pred / GT):} & \\
& \texttt{[N][N][N][N][N][N] / [N][N][N][N][N][N]} & \\
& \texttt{Video Mask (pred / GT):} & \\
& \texttt{[C][M][N][N][M][C] / [C][M][N][N][M][C]} \\
& \textbf{\highlight{yellow!30}{there was}\highlight{orange!20}{ no} heat \highlight{Green!10}{so there was no sound}} & \textbf{33.3} \\
\midrule
Ground-truth & there is no air so there is no sound & -- \\
\midrule
\midrule
\multirow{5}{*}{ASR 5-best} &
    it is the same & 80.0  \\
    & at the same time . & 100.0 \\
    & at the same time & 100.0 \\
    & which again opens\highlight{orange!20}{ the elements} & 60.0 \\
    & it is the same . & 100.0 \\
    \midrule
\multirow{5}{*}{VSR 5-best} &
    and it opens your eyes & 100.0 \\
    & \highlight{yellow!30}{it opens to} the enemies & 60.0 \\
    & \highlight{yellow!30}{it opens to} the animation & 60.0 \\
    & and \highlight{yellow!30}{it opens to} the enemies & 80.0 \\
    & and \highlight{yellow!30}{it opens to} the animation & 80.0 \\
\midrule
\textbf{DualHyp output} & \textbf{\highlight{yellow!30}{it opens to} the east} & \textbf{60.0} \\
\midrule
\multirow{5}{*}{\textbf{RelPrompt output}} & \texttt{Audio Mask (pred / GT):} & \\
& \texttt{[N][N][N][C][C] / [M][N][N][C][C]} & \\
& \texttt{Video Mask (pred / GT):} & \\
& \texttt{[C][N][N][N][C] / [C][M][N][N][C]} \\
& \textbf{\highlight{yellow!30}{it opens to}\highlight{orange!20}{ the elements}} & \textbf{40.0} \\
\midrule
Ground-truth & again open to the elements & -- \\
\midrule
\midrule
\multirow{5}{*}{ASR 5-best} &
    \highlight{yellow!30}{like one hundreds of one thousands of people}\highlight{orange!20}{ do every year} & 0.0  \\
    & like one hundreds or one thousands of people\highlight{orange!20}{ do every year} & 9.1 \\
    & one hundreds of one thousands of people do every year & 9.1 \\
    & \highlight{yellow!30}{like one hundreds of one thousands of people}\highlight{orange!20}{ do every year} . & 9.1 \\
    & like one hundreds and one thousands of people\highlight{orange!20}{ do every year} & 9.1 \\
    \midrule
\multirow{5}{*}{VSR 5-best} &
    \highlight{yellow!30}{like one hundreds of one thousands of people} or so every & 27.3 \\
    & \highlight{yellow!30}{like one hundreds of one thousands of people} or so every year & 18.2 \\
    & \highlight{yellow!30}{like one hundreds of one thousands of people} or so often & 27.3 \\
    & \highlight{yellow!30}{like one hundreds of one thousands of people} do every & 9.1 \\
    & \highlight{yellow!30}{like one hundreds of one thousands of people} or so whoever & 27.3 \\
\midrule
\textbf{DualHyp output} &
\textbf{like one hundreds or one thousands of people\highlight{orange!20}{  do every year}} & \textbf{9.1} \\
\midrule
\multirow{5}{*}{\textbf{RelPrompt output}} & \texttt{Audio Mask (pred / GT):} & \\
& \texttt{[C][N][N][N][N][C][C] / [C][N][N][N][N][C][C]} & \\
& \texttt{Video Mask (pred / GT):} & \\
& \texttt{[C][C][C][C][N][N][C] / [C][C][C][N][N][N][N]} \\
& \textbf{\highlight{yellow!30}{like one hundreds of one thousands of people}\highlight{orange!20}{ do every year}} & \textbf{0.0} \\
\midrule
Ground-truth & like one hundreds of one thousands of people do every year & -- \\
\bottomrule
\end{tabularx}
}
\caption{Qualitative examples of successful corrections by DualHyp with RelPrompt. These cases show how RelPrompt improves upon the baseline DualHyp by leveraging the predicted reliability masks (\texttt{pred}) to trust or discard certain parts of hypotheses from the ASR and VSR streams. Ground-truth masks (\texttt{GT}) are also shown for comparison.}
\label{tab:additional_qual_analysis}
\end{table*}

\subsection{Success Case}
As a supplement to the cases presented in Table~\ref{tab:dualhyp_case}, Table~\ref{tab:successful_cases} provides further qualitative examples that illustrate the successful mechanisms of our DualHyp framework.
These successes can be categorized into two main patterns.

The first pattern, \textit{Multimodal Fragment Composition}, involves the model's ability to recover correct transcriptions by leveraging complementary fragments from both ASR and VSR hypotheses.
This can be seen when the framework fuses the beginning of an ASR hypothesis and the end of a VSR hypothesis as in the first case (\ie, combining \texttt{which upset some...} from ASR and \texttt{...female residents} from VSR), or vice versa as in the second case (\ie, \texttt{so rather than...} and \texttt{relying on...}).
The compositional fusion of correct sub-sequences from noisy ASR and VSR inputs highlights the DualHyp's robustness by leveraging a generative correction ability of LLM.

The second pattern is \textit{Dominant Modality Refinement}, where the model identifies and grounds the prediction in the more reliable modality, even when that modality's best hypothesis is not perfect.
This is evident in \texttt{the armed forces} and \texttt{already in the states} cases, where the model primarily refines ASR's strong-but-flawed hypotheses while disregarding the less plausible VSR candidates.
These cases highlight that providing the LLM with separate, modality-specific hypotheses is a more effective correction strategy than relying on a single or early-fused representation, as it allows the model to reason over distinct evidences.

\subsection{Failure Case}
Following the successful cases, we also present and analyze several typical failures, which often occur when both modalities provide highly ambiguous information.
As illustrated in Table~\ref{tab:failure_cases}, these failures can be categorized into two primary patterns.

The first failure pattern is \textit{Over-reliance on Plausible but Inaccurate Hypotheses}, where LLMs are misled by a semantically incorrect candidate from one modality.
In the \texttt{probably in the world} example, the LLM disregards the partially correct ASR hypotheses and instead adopts the coherent but entirely wrong VSR hypothesis.
Second example shows that the model favors the plausible but incorrect verb \texttt{to wait} from VSR candidates, although the correct verb \texttt{to win} is also present in the other VSR hypotheses. 
These cases show that the ambiguity between hypotheses can make the LLM confuse and incorrectly prioritize a plausible but wrong candidate.
The over-reliance issue is a common drawback in all GER frameworks but can be mitigated to some extent by leveraging our RelPrompt, as shown in Figure~\ref{fig:qual_analysis}.

The second failure pattern involves \textit{Hallucination and Semantic Association Errors}, where the LLM generates words that are not present in any of the provided hypotheses.
This often occurs when the model is biased towards a specific keyword and generates a semantically related but incorrect term, as seen in the \texttt{end of november} example where it generates \texttt{december} out of nowhere.
In the last case, the model's strong prior knowledge can override direct evidence, misinterpreting \texttt{bat hotel} as \texttt{bistro}.
This reveals the fundamental duality of leveraging the LLM's internal knowledge for GER, where context-aware corrections produce not only useful generative revisions but also factually incorrect hallucinations, suggesting a potential direction for future research on controlling this mechanism.

\begin{table*}[!t]
    \centering
    \small
    \addtolength{\tabcolsep}{-2pt}
    \resizebox{\textwidth}{!}{
    \begin{tabular}{lcc|ccccc|ccccc}
        \toprule
        &&& \multicolumn{5}{c|}{\textbf{Object occlusion (50\%)}} & \multicolumn{5}{c}{\textbf{Speech noise (SNR = 0\,dB)}} \\
        \textbf{Method} & \textbf{+\,LRS3} & \textbf{+\,HR} & \textbf{Babble} & \textbf{Speech} & \textbf{Music} & \textbf{Natural} & \textbf{Overall} & \textbf{Object} & \textbf{Hands} & \textbf{Pixelate} & \textbf{Blur} & \textbf{Overall} \\
        \midrule
        \multirow{3}{*}{GER} & \xmark & \xmark & 39.3 & 34.4 & 11.5 & 13.2 & 24.6 & - & - & - & - & 23.9 \\
         & \cmark & \xmark & 39.1 & 34.3 & 11.7 & 12.8 & 24.5 & - & - & - & - & 23.9 \\
         & \xmark & \cmark & 39.2 & 34.1 & 11.7 & 13.1 & 24.5 & - & - & - & - & 25.6 \\
        \midrule
        \multirow{3}{*}{\textbf{DualHyp}} & \xmark & \xmark & 21.6 & 17.9 & 8.1 & 9.3 & 14.2 & 12.0 & 11.8 & 12.7 & 11.1 & 11.9 \\
         & \cmark & \xmark & 21.0 & 17.9 & 7.8 & 8.7 & 13.9 & 12.7 & 11.4 & 12.3 & 11.2 & 11.9 \\
         & \xmark & \cmark & 21.9 & 17.7 & \textbf{7.7} & \textbf{8.0} & 13.8 & 12.1 & 11.0 & 11.5 & 10.5 & 11.3 \\
         \midrule
        \multirow{3}{*}{\thead[l]{\textbf{DualHyp}\\\textbf{+\,RelPrompt}}} & \xmark & \xmark & 20.4 & 16.0 & 8.0 & 8.2 & 13.2 & 11.9 & 11.0 & 11.9 & 10.2 & 11.3 \\
         & \cmark & \xmark & \textbf{19.5} & 15.4 & 7.8 & 8.5 & \textbf{12.8} & 11.7 & 11.0 & 11.9 & \textbf{9.6} & 11.1 \\
         & \xmark & \cmark & 20.1 & \textbf{15.1} & \textbf{7.7} & 8.3 & \textbf{12.8} & \textbf{10.5} & \textbf{9.4} & \textbf{10.9} & 9.7 & \textbf{10.1} \\
        \bottomrule
    \end{tabular}
    }
    \caption{The effect of dataset integration (+\,LRS3) and high-resource (+\,HR) training.}
    \label{tab:high_resource}
\end{table*}

\begin{figure*}[!t]
    \centering
    \includegraphics[width=.48\linewidth]{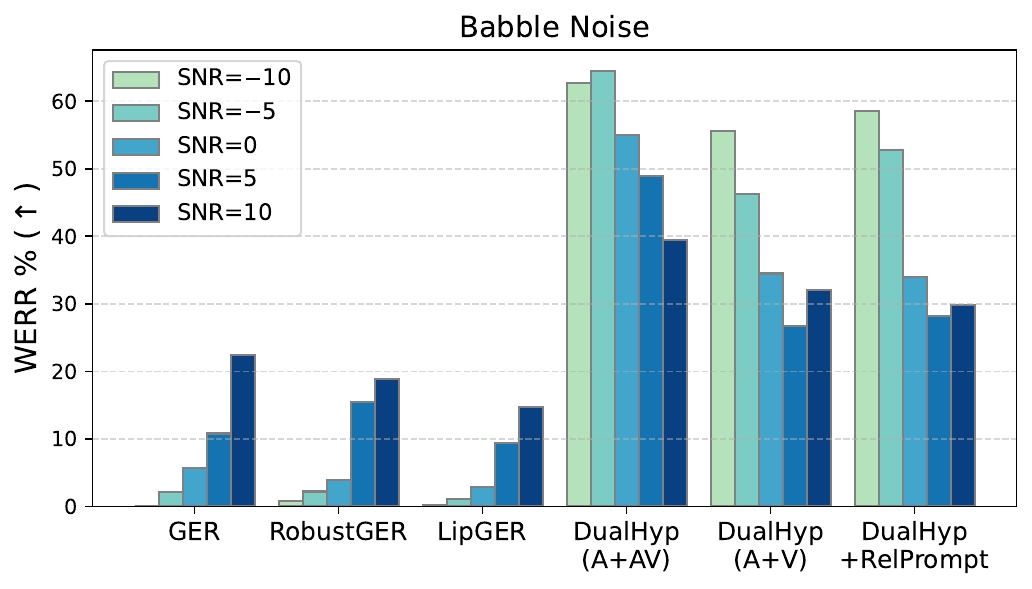}
    \includegraphics[width=.48\linewidth]{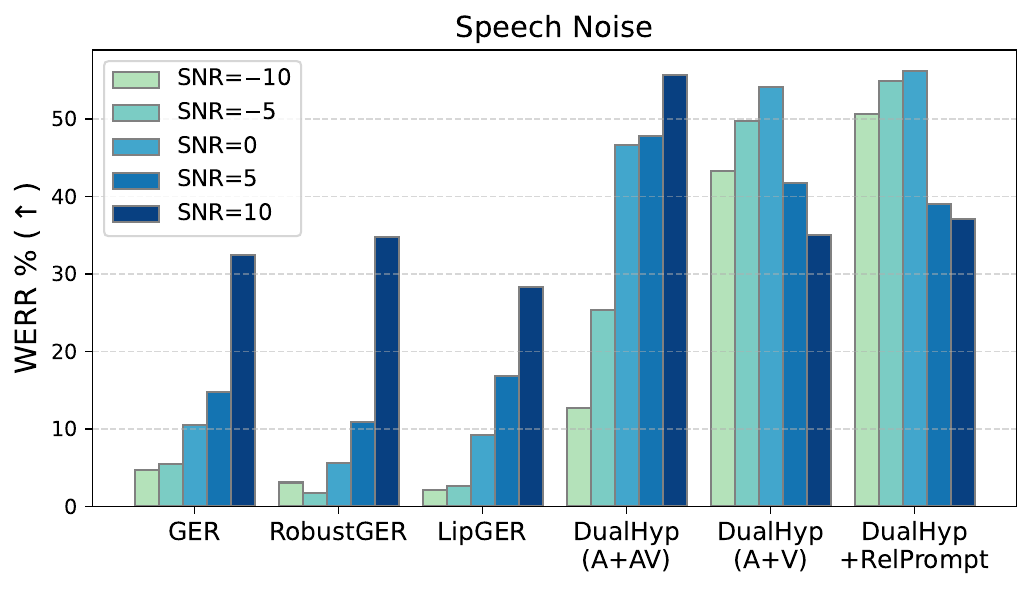}

    \includegraphics[width=.48\linewidth]{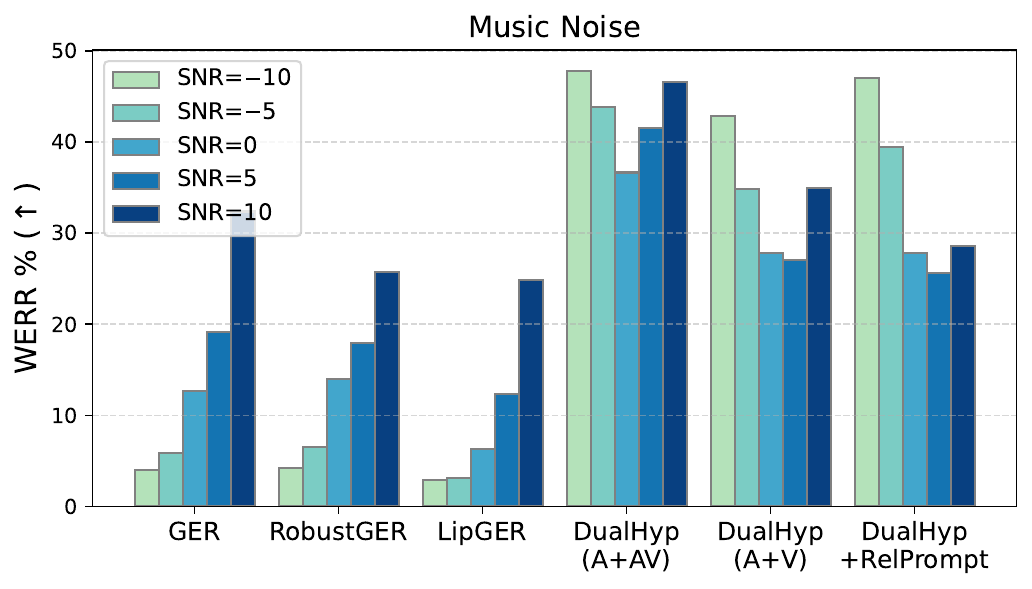}
    \includegraphics[width=.48\linewidth]{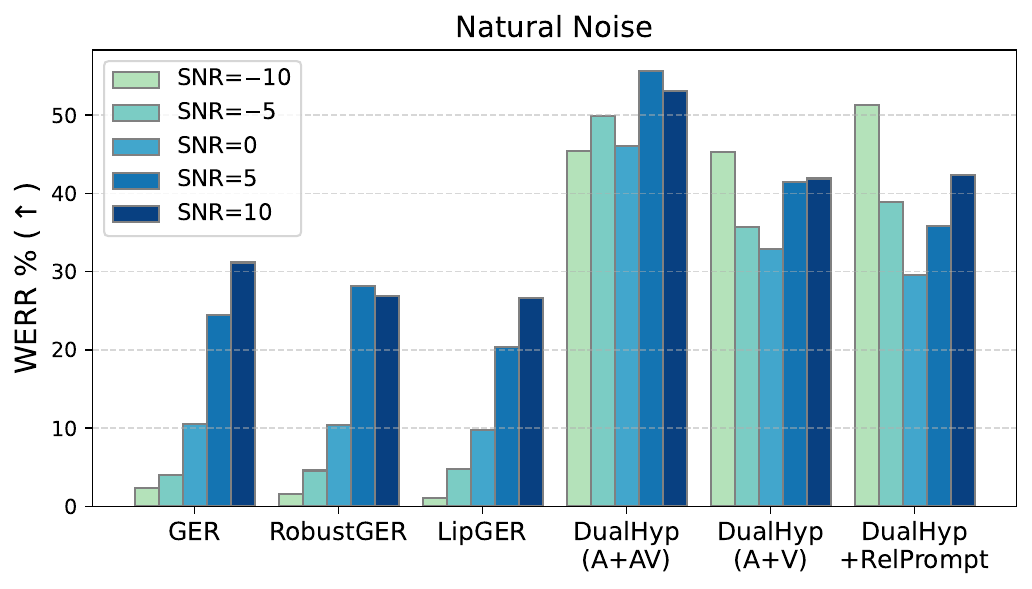}
    \vspace{-5pt}
    \caption{Word error rate reduction (WERR) at different audio SNRs, under diverse types of noise. Higher WERR indicates greater improvement over the Whisper ASR baseline. 
    The experimental setup is identical to Table~\ref{tab:main_lrs2}\textcolor{Red}{a}.
    }
    \label{fig:additional_snr_analysis}
    \vspace{10pt}
\end{figure*}

\begin{table*}[!t]
    \centering
    \small
    \addtolength{\tabcolsep}{-2.5pt}
    \resizebox{\textwidth}{!}{
    \begin{tabular}{lc|llllllll}
    \toprule
    \textbf{Method} & \textbf{Input} & \textbf{Arabic} & \textbf{German} & \textbf{Greek} & \textbf{Spanish} & \textbf{French} & \textbf{Italian} & \textbf{Portuguese} & \textbf{Russian} \\
    \midrule
    Whisper-large-v3 & A & \textbf{91.7} & \textbf{55.7} & \textbf{54.4} & 49.6 & \textbf{46.8} & 52.3 & 52.7 & \textbf{50.9} \\
    mAV-HuBERT & V & 102.0$_{(+11\%)}$ & 96.6$_{(+73\%)}$ & 87.1$_{(+60\%)}$ & 70.5$_{(+42\%)}$ & 81.7$_{(+75\%)}$ & 73.7$_{(+41\%)}$ & 74.1$_{(+41\%)}$ & 80.9$_{(+59\%)}$ \\
    GER & A & 96.9$_{(+6\%)}$ & 56.2$_{(+1\%)}$ & 57.7$_{(+6\%)}$ & 50.6$_{(+2\%)}$ & 47.8$_{(+2\%)}$ & 58.5$_{(+12\%)}$ & 52.3$_{(-1\%)}$ & 54.1$_{(+6\%)}$ \\
    \textbf{DualHyp (ours)} & A\,+\,V & 106.8$_{(+16\%)}$ & 100.4$_{(+80\%)}$ & 77.3$_{(+42\%)}$ & \textbf{47.3}$_{(-5\%)}$ & 47.9$_{(+2\%)}$ & \textbf{47.2}$_{(-10\%)}$ & \textbf{49.0}$_{(-7\%)}$ & 58.9$_{(+16\%)}$ \\
    \bottomrule
    \end{tabular}
    }
    \caption{WER (\%) comparison with multilingual babble noise (SNR\,=\,0\,dB) on the MuAViC dataset.
    Subscript values of mAV-HuBERT indicate the relative WER increase compared to Whisper-large-v3.}
    \label{tab:muavic_full}
\end{table*}

\subsection{RelPrompt}
Table~\ref{tab:additional_qual_analysis} provides the qualitative examples that demonstrate how RelPrompt successfully corrects errors for the cases where baseline DualHyp framework fails. In the first example (\texttt{there is no air...}), DualHyp is misled by entirely incorrect ASR hypotheses (\texttt{your baby of...asked}). RelPrompt, in contrast, uses its predicted audio reliability tokens (all \texttt{[N]}) and rather clean video reliability tokens (\texttt{[C]}) to correctly identify the audio stream as unreliable, allowing it to pivot to the more accurate VSR hypotheses for a much better result. 

In the second and third examples, the reliability masks guide the model to capitalize on the structure from the cleaner VSR stream at the beginning of the utterance, while correctly extracting a more accurate key phrase (\ie, \texttt{the elements} and \texttt{do every year}) from the ASR stream to form the ending.
The baseline DualHyp method, lacking this guidance, is confused by the conflicting signals and produces errors by incorporating some flawed hypotheses.
These cases demonstrate how the explicit reliability signals empower the model to intelligently arbitrate between hypotheses at a sub-sentence level, composing the final output from the most reliable fragments of each modality.

\section{Additional Results}
\label{app:additional_results}

\subsection{High-Resource Training}
\label{subsec:high-reource_training}

We investigate how our framework scales with additional training data by augmenting the main LRS2 training set (29 hours) with either the larger LRS3 dataset (59 hours) or a high-resource LRS2 pretraining set (HR, 195 hours). The results in Table~\ref{tab:high_resource} show that GER fails to benefit from more data, showing no to adverse impact on the performance. In contrast, our DualHyp frameworks consistently improve with larger training sets. The best performance is achieved by DualHyp\,+\,RelPrompt when trained with the high-resource data, reaching 12.8\% overall WER on audio corruptions and 10.1\% overall WER on visual corruptions. This indicates that while the bottleneck of single-stream GER is not readily resolved by scaling data, our compositional framework has the capacity to effectively leverage more data to enhance its robustness.

\begin{table*}[!t]
  \centering
  \small
  \resizebox{\textwidth}{!}{
  \begin{tabular}{lc|llll|l}
    \toprule
    \textbf{Method} & \textbf{Input} & \textbf{Babble (B)} & \textbf{Speech (S)} & \textbf{Music (M)} & \textbf{Natural (N)} & \textbf{Overall (O)} \\
    \midrule
    \color{gray}\textit{ASR oracle $o_{nb}$\,/\,$o_{cp}$} & \color{gray}A & \color{gray}26.9 / 23.2 & \color{gray}19.6 / 13.5 & \color{gray}4.5 / 3.7 & \color{gray}5.0 / 4.6 & \color{gray}14.0 / 11.2 \\
    \color{gray}{\textit{ASR\,+\,VSR oracle $o_{nb}$\,/\,$o_{cp}$}} & \color{gray}A\,+\,V & \color{gray}7.9 / 5.8 & \color{gray}5.6 / 3.5 & \color{gray}1.6 / 1.0 & \color{gray}1.9 / 1.3 & \color{gray}4.2 / 2.9 \\
    \midrule
    Whisper-large-v3 & A & 32.6 & 34.5 & 7.3 & 7.8 & 20.6 \\
    BRAVEn-large & V & - & - & - & - & 31.9$_{(+54.9\%)}$ \\
    GER~\cite{chen2023hyporadise} & A & 32.4$_{(-0.6\%)}$ & 35.4$_{(+2.6\%)}$ & 7.6$_{(+4.1\%)}$ & 8.0$_{(+2.6\%)}$ & 20.9$_{(+1.5\%)}$ \\
    RobustGER~\cite{hu2024robustger} & A & 32.5$_{(-0.3\%)}$ & 36.0$_{(+4.3\%)}$ & 7.7$_{(+5.5\%)}$ & 8.1$_{(+3.8\%)}$ & 21.1$_{(+2.4\%)}$ \\
    LipGER~\cite{ghosh2024lipger} & AV & 32.4$_{(-0.6\%)}$ & 34.4$_{(-0.3\%)}$ & 7.6$_{(+4.1\%)}$ & 8.1$_{(+3.8\%)}$ & 20.6$_{(-0.0\%)}$ \\
    GER w/ Auto-AVSR$^\dagger$ & AV & 17.9$_{(-45.1\%)}$ & 45.6$_{(+32.2\%)}$ & 14.2$_{(+94.5\%)}$ & 11.0$_{(+41.0\%)}$ & 22.2$_{(+7.8\%)}$ \\
    \midrule
    \textbf{DualHyp (ours)} & A\,+\,V & 16.3$_{(-50.0\%)}$ & 18.2$_{(-47.2\%)}$ & \textbf{5.6}$_{(-23.3\%)}$ & {5.5}$_{(-29.5\%)}$ & 11.4$_{(-44.7\%)}$ \\
    \textbf{+\,RelPrompt (ours)} & A\,+\,V & \textbf{14.9}$_{(-54.3\%)}$ & \textbf{16.2}$_{(-53.0\%)}$ & {5.7}$_{(-21.9\%)}$ & \textbf{5.1}$_{(-34.6\%)}$ & \textbf{10.5}$_{(-49.0\%)}$ \\
    \bottomrule
    \addlinespace[2pt]
    \multicolumn{7}{c}{\textbf{(a) Audio: random noise [-10, 10]\,dB, Video: 50\% segment occluded with object}} \\
    \addlinespace[8pt]
    \toprule
    \textbf{Method} & \textbf{Input} & \textbf{Object} & \textbf{Hands} & \textbf{Pixelate} & \textbf{Blur} & \textbf{Overall} \\
    \midrule
    \color{gray}\textit{ASR oracle $o_{nb}$\,/\,$o_{cp}$} & \color{gray}A & \color{gray}- & \color{gray}- & \color{gray}- & \color{gray}- & \color{gray}8.3 / 5.3 \\
    \color{gray}{\textit{ASR\,+\,VSR oracle $o_{nb}$\,/\,$o_{cp}$}} & \color{gray}A\,+\,V & \color{gray}3.2 / 1.8 & \color{gray}3.0 / 1.6 & \color{gray}3.0 / 1.5 & \color{gray}2.5 / 1.4 & \color{gray}2.9 / 1.6 \\
    \midrule
    Whisper-large-v3 & A & 23.8 & 23.8 & 23.8 & 23.8 & 23.8 \\
    BRAVEn-large & V & 31.9$_{(+34.0\%)}$ & 30.8$_{(+29.4\%)}$ & 29.5$_{(+23.9\%)}$ & 23.8$_{(-0.0\%)}$ & 29.0$_{(+21.8\%)}$ \\
    GER~\cite{chen2023hyporadise} & A & - & - & - & - & 26.0$_{(+9.2\%)}$ \\
    RobustGER~\cite{hu2024robustger} & A & - & - & - & - & 27.1$_{(+13.9\%)}$ \\
    LipGER~\cite{ghosh2024lipger} & AV & 26.2$_{(+10.1\%)}$ & 26.0$_{(+9.2\%)}$ & 25.9$_{(+8.8\%)}$ & 26.0$_{(+9.2\%)}$ & 26.0$_{(+9.2\%)}$ \\
    GER w/ Auto-AVSR$^\dagger$ & AV & 47.5$_{(+99.6\%)}$ & 44.2$_{(+85.7\%)}$ & 42.0$_{(+76.5\%)}$ & 38.2$_{(+60.5\%)}$ & 43.0$_{(+80.7\%)}$ \\
    \midrule
    \textbf{DualHyp (ours)} & A\,+\,V & {12.2}$_{(-48.7\%)}$ & 10.9$_{(-54.2\%)}$ & 10.8$_{(-54.6\%)}$ & {9.6}$_{(-59.7\%)}$ & 10.9$_{(-54.2\%)}$ \\
    \textbf{+\,RelPrompt (ours)} & A\,+\,V & \textbf{11.0}$_{(-53.8\%)}$ & \textbf{10.5}$_{(-55.9\%)}$ & \textbf{10.1}$_{(-57.6\%)}$ & \textbf{8.8}$_{(-63.0\%)}$ & \textbf{10.1}$_{(-57.6\%)}$ \\
    \bottomrule
    \addlinespace[2pt]
    \multicolumn{7}{c}{\textbf{(b) Audio: speech noise 0\,dB, Video: random segment corrupted}}
  \end{tabular}
  }
  \vspace{-5pt}
  \caption{
  WER\% ($\downarrow$) results on the LRS3 test set under joint audio-visual corruption. (a) Performance across varying audio noise types, with a fixed visual corruption (50\% segment occluded by an object). (b) Performance across varying visual corruption types, with a fixed audio corruption (0\,dB speech noise). We also show the relative WER reduction in parentheses compared to the Whisper-large-v3 ASR baseline. All the ASR and VSR heads are Whisper-large-v3 and BRAVEn-large, respectively. $^\dagger$: We implement a GER model using hypotheses generated from an early-fusion approach, Auto-AVSR~\cite{ma2023auto}, which has been trained on LRS3 with babble noise.
  }
  \label{tab:main_lrs3}
\end{table*}

\subsection{SNR-wise WER Improvement}

Figure~\ref{fig:additional_snr_analysis} provides the entire result of Figure~\ref{fig:snr_analysis} with WERR across all four audio noise types. 
Across all conditions, the single-stream baselines show that WERR is proportional to the audio quality, providing significant gains only at high SNRs. 
The DualHyp (A\,+\,AV) variant also illustrates this principle; it achieves a high WERR on familiar babble noise but does not show such strong correction capabilities on speech noise, especially at low SNRs.
This demonstrates a key limitation of the early-fusion AVSR head: since it is affected by the audio corruption, it may fail to provide a truly independent and useful signal for error correction.
In contrast, our DualHyp frameworks demonstrate superior robustness by maintaining high WERR even at very low SNRs, effectively leveraging the visual stream when the audio is most corrupted.

\subsection{MuAViC Results}

Table~\ref{tab:muavic_full} presents our full multilingual results on the MuAViC dataset~\cite{anwar2023muavic}.
We observe that standard GER shows limited effectiveness across all languages, suggesting inherent difficulty of error correction in non-English contexts, even when the LLM itself is multilingual.
Our DualHyp framework is designed to aid this reasoning by providing the LLM with more comprehensive evidence from both ASR and VSR streams, achieving performance improvements in three languages.

However, our results reveal that a large disparity between the ASR and VSR quality can exacerbate the LLM's inherent weakness in multilingual reasoning.
While for the languages where DualHyp succeeds, the VSR head maintains a relatively consistent performance gap around 40\% higher than the ASR baseline, for the languages where DualHyp underperforms (\eg, Greek), this gap widens significantly to over 60\%.
Meanwhile, for Arabic, the hypotheses from both modalities are of exceptionally poor quality (>90\% WER), leaving the LLM with no useful source to compose.

\begin{table}[!t]
    \centering
    \small
    \vspace{5pt}
    \addtolength{\tabcolsep}{-1pt}
    \resizebox{\columnwidth}{!}{
    \begin{tabular}{lcc|cccc|c}
        \toprule
        \textbf{Method} & \!\textbf{Input}\! & \!\!\textbf{\# hyps}\!\! & \textbf{B} & \textbf{S} & \textbf{M} & \textbf{N} & \textbf{O} \\
        \midrule
        \multirow{3}{*}{GER} & A & 5 & 32.4 & 35.4 & 7.6 & 8.0 & 20.9 \\
         & AV & 5 & 17.9 & 45.6 & 14.2 & 11.0 & 22.2 \\
         & AV & 10 & 18.3 & 44.8 & 14.1 & 10.6 & 21.9  \\
        \midrule
        \multirow{2}{*}{\textbf{DualHyp}} & \!\!A\,+\,AV\!\! & 10 & 19.3 & 34.8 & 6.0 & {5.4} & 16.4 \\
        & A\,+\,V & 10 & {16.3} & {18.2} & \textbf{5.6} & 5.5 & {11.4} \\
        \midrule
        \textbf{DualHyp} & \!\!A\,+\,AV\!\! & 10 & 19.0 & 32.9 & 6.2 & 6.9 & 16.3 \\
        \textbf{+\,RelPrompt} & A\,+\,V & 10 & \textbf{14.9} & \textbf{16.2} & {5.7} & \textbf{5.1} & \textbf{10.5} \\
        \bottomrule
    \end{tabular}
    }
    \caption{
    WER (\%) comparison of different hypotheses from single-stream (GER) and dual-stream (DualHyp) generation heads, on the LRS3 dataset. 
    The corruption strategy follows Table\,\ref{tab:main_lrs3}\textcolor{Red}{a}.}
    \label{tab:comparison_avsr_lrs3}
    \vspace{-5pt}
\end{table}

\subsection{LRS3 Results}
\label{app:lrs3_result}

Similar to Table~\ref{tab:main_lrs2} and \ref{tab:comparison_avsr} in the main paper, Table~\ref{tab:main_lrs3} and \ref{tab:comparison_avsr_lrs3} respectively present additional results on the LRS3 dataset~\citep{afouras2018lrs3} to demonstrate the generalizability of our findings. A key difference from the LRS2 experiments is that our VSR head, BRAVEn-large, has also been fine-tuned on LRS3, making it a much stronger, in-domain model supporting the ASR stream. This serves to amplify the benefits of our dual-stream approach.

As shown in Table~\ref{tab:main_lrs3}, our DualHyp\,+\,RelPrompt framework achieves an overall WER of 10.5\% on audio corruptions and 10.1\% on visual corruptions. The performance gap between our method and GER w/ Auto-AVSR is even larger than on LRS2 (also refer to Table~\ref{tab:comparison_avsr_lrs3}), confirming that as the quality of the independent VSR head improves, the advantage of our language-space fusion becomes more pronounced.
We also observe that on LRS3, the ASR hypotheses, while coherent, are often homogeneous and contain similar errors across the $N$-best list. Our DualHyp approach is particularly effective in this case, as the independent VSR hypotheses provide the diversity to break out of the ASR's error patterns.

\section{LLM Usage}
We acknowledge the use of LLM in this paper as a writing assistant, with its role strictly limited to supporting tasks such as paraphrasing, grammar check, and improving clarity. All conceptual contributions, method development, experimental design, and analysis are done without using LLMs or AI Agents.

\end{document}